\documentclass[twocolumn,traditabstract,longauth]{aa}
\usepackage{amssymb}
\usepackage{mathptmx}
\usepackage{natbib}
\bibpunct{(}{)}{;}{a}{}{,}
\usepackage[]{graphicx}
\usepackage{lscape}
\usepackage{txfonts}
\usepackage{url}
\defcitealias{canameras15}{C15}
\defcitealias{canameras17b}{C17a}
\begin{document}

\title{\textit{Planck}'s dusty GEMS. V. Molecular wind and clump stability in a 
  strongly lensed star-forming galaxy at \textit{z} = 2.2 \thanks{Based on data
    obtained with the following programs: IRAM Plateau de Bure Interferometer
    program ID: X0AE. Canada-France-Hawaii Telescope program ID: 14AF06. 
    Submillimeter Array program ID: 2013B-S050. {\it Spitzer} Space Telescope 
    program ID: 10010.}}

\author{R.~Ca\~nameras\inst{1,2,3}, N.~P.~H.~Nesvadba\inst{2}, M.~Limousin\inst{4}, 
H.~Dole\inst{2}, R.~Kneissl\inst{5,6}, S.~Koenig\inst{7}, E.~Le~Floc'h\inst{8}, 
G.~Petitpas\inst{9}, and~D.~Scott\inst{10}}

\institute{
Dark Cosmology Centre, Niels Bohr Institute, University of Copenhagen, 
Juliane Maries Vej 30, DK-2100 Copenhagen, Denmark
\and
Institut d'Astrophysique Spatiale, CNRS, Univ. Paris-Sud, Universit\'e 
Paris-Saclay, B\^at. 121, 91405 Orsay France 
\and 
email: {\tt canameras@dark-cosmology.dk}
\and
Aix Marseille Universit\'e, CNRS, CNES, LAM, Marseille, France
\and
European Southern Observatory, ESO Vitacura, Alonso de Cordova 3107, 
Vitacura, Casilla 19001, Santiago, Chile
\and
Atacama Large Millimeter/submillimeter Array, ALMA Santiago Central 
Offices, Alonso de Cordova 3107, Vitacura, Casilla 763-0355, Santiago, 
Chile
\and
Chalmers University of Technology, Onsala Space Observatory, Onsala, 
Sweden
\and
Laboratoire AIM, CEA/DSM/IRFU, CNRS, Universit\'e Paris-Diderot, B\^at. 
709, 91191 Gif-sur-Yvette, France
\and
Harvard-Smithsonian Center for Astrophysics, Cambridge, MA 02138, USA
\and
Department of Physics and Astronomy, University of British Columbia, 
6224 Agricultural Road, Vancouver, 6658 British Columbia, Canada}

\titlerunning{Molecular wind and clump stability in the Emerald at \textit{z} = 2.2}

\authorrunning{R. Ca\~nameras et al.}  \date{Received / Accepted }

\abstract{We report the discovery of a molecular wind signature from a 
massive intensely star-forming clump of a few $10^9$~M$_{\odot}$, in the 
strongly gravitationally lensed submillimeter galaxy ``the Emerald'' 
(PLCK\_G165.7+49.0) at $z=2.236$. The Emerald is amongst the brightest 
high-redshift galaxies on the submillimeter sky, and was initially 
discovered with the \textit{Planck} satellite. The system contains two 
magnificient structures with projected lengths of 28.5\arcsec\ and 
21\arcsec\ formed by multiple, near-infrared arcs, falling behind a 
massive galaxy cluster at $z=0.35$, as well as an adjacent filament that 
has so far escaped discovery in other wavebands. 
We used {\it HST}/WFC3 and CFHT optical and near-infrared imaging together
with IRAM and SMA interferometry of the CO(4--3) line and 850~$\mu$m
dust emission to characterize the foreground lensing mass distribution, 
construct a lens model with {\sc Lenstool}, and calculate 
gravitational magnification factors between 20 and 50 in most of the 
source. The majority of the star formation takes place within two 
massive star-forming clumps which are marginally gravitationally bound
and embedded in a $9 \times 10^{10}$ M$_{\odot}$, fragmented disk with
20\% gas fraction. The stellar continuum morphology is much smoother
and also well resolved perpendicular to the magnification axis. One of
the clumps shows a pronounced blue wing in the CO(4--3) line profile,
which we interpret as a wind signature. The mass outflow rates are high 
enough for us to suspect that the clump might become unbound within a 
few tens of Myr, unless the outflowing gas can be replenished by gas
accretion from the surrounding disk. The velocity offset of --200~km
s$^{-1}$ is above the escape velocity of the clump, but not that of the 
galaxy overall, suggesting that much of this material might ultimately 
rain back onto the galaxy and contribute to fueling subsequent star
formation.}


\maketitle

\section{Introduction}
\label{sec:introduction}

Rapid, intense star formation that occurs in dusty star-forming galaxies
at $z \sim 1$--4 \citep{casey14} is expected to dominate the cosmic
star-formation rate density at these epochs \citep[e.g.,][]{dole06} and 
corresponds to the early growth phase of giant ellipticals seen in 
high-density regions of the local Universe 
\citep[e.g.,][]{lilly99,swinbank06}. Intense star formation is sustained 
for timescales up to a few hundred Myr in these high-redshift galaxies. 
Many recent studies propose that the global properties of the molecular 
gas reservoirs, including gas fractions, determine whether these galaxies 
will fall on the high-redshift main-sequence of star formation 
\citep{daddi10,genzel10}, or be in the starburst mode or the transition 
regime \citep[e.g.,][]{tacconi13,dessauges15,lee17,canameras17b}. 
Variations in star-formation efficiency might also play a role in this 
regard \citep[e.g.,][]{hodge15,genzel15,usero15}.

The role of local mechanisms such as star-formation feedback, winds
and turbulence in shaping the interstellar medium of these galaxies,
in regulating their star-formation activity and in triggering their 
major growth phase is still a matter of active debate. For instance,
large-scale outflows are a major component of galaxy evolution models
\citep[e.g.,][]{hopkins06}, since they can affect and even quench
star formation within the hosts by expelling the gas to the  
circumgalactic medium. Molecular outflows are ubiquitous in nearby
ULIRGs \citep[e.g.,][]{weiss99,sturm11,cicone14,veilleux13} and are 
attributed to either feedback from star formation, or from a central 
AGN or both. At high redshift, ouflows have been almost exclusively 
detected in ionized gas \citep[e.g.,][]{barger99,nesvadba07,newman12}, 
so it remains unclear how intense winds affect the molecular gas 
reservoirs. This emphasizes the need to increase the number of 
high-redshift dusty star-forming galaxies with measurements of the local 
gas kinematics and of the stellar mass, gas mass and star-formation 
surface densities \citep[e.g.,][]{hatsukade15}. This must be done down 
to typical disk-fragmentation scales \citep{toomre64,escala08}, in 
order to probe local energy injection from a range of feedback processes 
and to characterize the resolved Schmidt-Kennicutt law 
\citep[e.g.,][]{swinbank11}.

Star formation in more than half of the high-redshift dust-obscured
galaxies appears to occur within massive giant star-forming clumps of
$10^7$--$10^9$~M$_{\odot}$ and size of about 1~kpc or less
\citep[e.g.,][]{elmegreen05}, embedded within more diffuse disks
\citep{swinbank11}. These clumpy structures were originally identified
in rest-frame UV and optical studies, and their properties play a central
role in the overall evolution of the host galaxies
\citep[e.g.,][]{tacconi13,mayer16,dessauges17,cava18}. Clumps with
sufficiently long lifetimes of a few 100~Myr survive the feedback from
young stellar populations and could migrate inward to form the central
bulge of galaxies \citep[e.g.,][]{ceverino10,bournaud14}. However, 
processes such as clump mergers, gas accretion, dynamical interactions 
within the disk and star-formation feedback could lead to their dissolution 
on much shorter timescales \citep[e.g.,][]{tamburello15}. A better
understanding of high-redshift star-formation process therefore requires
us to resolve the gas, dust and stellar properties of these clumps. This
is best achieved in very strongly gravitationally lensed galaxies for
which the lensing magnifications extend the resolution limits of current 
facilities and boost the apparent source brightness.

Here we present optical and near-infrared imaging and submillimeter and 
millimeter interferometry of PLCK\_G165.7+49.0, a strongly gravitationally 
lensed dusty star-forming galaxy at $z=2.236$. This source was identified as 
part of our \textit{Planck}'s Dusty Gravitationally Enhanced subMillimeter 
Sources (GEMS) follow-up program of 11 of the brightest high-redshift 
galaxies on the submillimeter sky discovered with the \textit{Planck} 
all-sky survey and \textit{Herschel} space observatory 
\citep[][]{dole15,canameras15,montier16}. PLCK\_G165.7+49.0 comprises a 
bright submillimeter arc, which we refer to as the ``Emerald'', near 
several other extended arcs falling behind a rich foreground environment, 
as recently discovered with CFHT imaging \citep{canameras15,canameras16}.

The luminous dusty starburst galaxy PLCK\_G165.7+49.0 has apparent 
far-infrared (FIR) luminosity of $\mu~L_{\rm FIR}=(1.0 \pm 0.1) \times 
10^{14}$~L$_{\odot}$ originating from $\mu~M_{\rm d}=(5.1 \pm 0.1) \times 
10^9$~M$_{\odot}$ of dust heated to a temperature of $T_{\rm d}=42.5 \pm 0.3$~K
\citep[][hereafter \citetalias{canameras15}]{canameras15}, where $\mu$ 
indicates the gravitational magnification factor. The far-infrared radio
correlation does not suggest the presence of a radio-loud AGN in this system, 
and photometric constraints from WISE at 22~$\mu$m, IRAS at 60 and 100~$\mu$m, 
and SPIRE at 250~$\mu$m do not suggest more than at most a few percent
AGN contamination to the overall FIR luminosity \citepalias{canameras15}. 
We also detected luminous CO(3--2) line emission with the wide-band 
heterodyne receiver EMIR on the 30-m telescope of IRAM, with an integrated 
flux of $\mu~I_{\rm CO}=25.4 \pm 0.3$~Jy~km~s$^{-1}$ and a line 
full-width-at-half-maximum (FWHM) of 580~km~s$^{-1}$ \citepalias{canameras15}.

To further characterize this source, we obtained CFHT and \textit{Spitzer}
optical and near-infrared imaging, as well as subarcsecond
submillimeter and millimeter interferometry of the dust and CO(4--3)
line emission with the Submillimeter Array (SMA) and the IRAM Plateau de
Bure Interferometer (PdBI). We used these data to characterize the 
foreground lensing potential, a hitherto unknown massive galaxy cluster at
$z=0.348$ with an adjacent filament, to calculate a strong lensing model 
with {\sc Lenstool}, and to characterize the gas kinematics and spatially
resolved dust and star formation properties of the background
source. To constrain the lensing model and characterize the stellar
components in the rest-frame UV, including their morphologies, we also
used recently obtained \textit{HST}/WFC3 imaging through the F110W and 
F160W filters, which are described in more detail in \citet{frye18}.

We present our analysis as follows. In 
Sect.~\ref{sec:observations} we describe our observations, data
reduction, and the construction of the photometric catalogs and
spectral imaging maps. In Sect.~\ref{sec:foreground}, we characterize 
the foreground structure with a three-way approach, by quantifying the 
local overdensity using an adaptive kernel density estimator, by 
identifying the red sequence of passively evolving member galaxies of 
the foreground structure, and by estimating photometric redshifts. In 
Sect.~\ref{sec:lenstool} we compute a strong lensing model based on the 
results of Sect.~\ref{sec:foreground} and the positions and brightnesses 
of foreground and background galaxies. In Sect.~\ref{sec:lemeraude} we 
characterize the intrinsic stellar, dust, gas and star formation 
properties of the Emerald. In Sect.~\ref{sec:gasenergetics}, we discuss 
the stability of the star-forming clumps, present the first detection of 
a molecular wind at high-redshift, and investigate whether it results 
from the kinetic energy and momentum injection from star formation. We 
then conclude with a summary in Sect.~\ref{sec:summary}.

Throughout the paper we have adopted the flat $\Lambda$CDM cosmology 
from \citet{planck16}, with $H_0 = 67.81$~km~s$^{-1}$~Mpc$^{-1}$,
$\Omega_{\rm M}$ = 0.308, and $\Omega_\Lambda = 1-\Omega_{\rm M}$. At the
redshift $z=2.236$ of PLCK\_G165.7+49.0, this corresponds to a luminosity
distance $d_{\rm L,bg} = 18.25$~Gpc, with a projected scale of 8.40~kpc
arcsec$^{-1}$. With the same cosmology, the luminosity distance of a
foreground source at $z=0.348$ is $d_{\rm L,fg} = 1.90$ Gpc. At that 
redshift, 5.07~kpc corresponds to 1\arcsec\ on the sky. All magnitudes 
are in the AB system.

\begin{table}
\centering
\begin{tabular}{lcccc}
\hline
\hline
ID & RA & Dec & $z_{\rm spec}$ & $\sigma_{\rm abs}$ \\
 & (J2000) & (J2000) & & [km s$^{-1}$] \\
\hline
G1 & 11:27:16.59 & +42:28:41.0 & 0.34788 $\pm$ 0.00007 & 323 $\pm$ 22 \\
G2 & 11:27:16.69 & +42:28:38.2 & 0.33767 $\pm$ 0.00004 & 307 $\pm$ 14 \\
G3 & 11:27:13.75 & +42:28:22.6 & 0.34770 $\pm$ 0.00012 & 271 $\pm$ 35 \\
\hline
\end{tabular}
\caption{Characteristics of three massive galaxies within the foreground structure 
of PLCK\_G165.7+49.0 taken from the 12th data release of the Sloan
Digital Sky Survey. The columns are: source name; right ascension and 
declination; spectroscopic redshift; and velocity dispersion measured 
from stellar absorption lines.}
\label{tab:sdssspectra}
\end{table}

\section{Observations and data reduction}
\label{sec:observations}

\subsection{Optical and near-infrared imaging}
\label{ssec:obs.optnir}

We obtained optical and near-infrared wide-field imaging of
PLCK\_G165.7+49.0 and surrounding sky with MEGACAM and WIRCAM on the
Canadia-France-Hawaii Telescope (CFHT) through the $r$-, $z$-, $J$-, and 
$K_{\rm s}$-band filters, and with the IRAC camera through the 3.6~$\mu$m and
4.5~$\mu$m filters on the \textit{Spitzer} Space Telescope.

At the CFHT, PLCK\_G165.7+49.0 was observed during several nights between
March and May 2014, as part of program 14AF06 (PI: Nesvadba). We
obtained a total of 40~min and 49~min of on-source observing time with
MEGACAM through the $r$- and $z$-band filters, respectively, and 93 and
52~min with WIRCAM through the $J$- and $K_{\rm s}$-band filters, respectively. 
The seeing was between 0.8 and 1.1\arcsec\ in the optical and the $J$-band,
and 0.7\arcsec\ in the $K_{\rm s}$-band. The near-infrared detectors were 
read out once every 10s in $J$, and once every 15s in $K_{\rm s}$.

Optical and near-infrared (NIR) images were bias and dark-frame subtracted, 
respectively, and all frames were flat-fielded before being released to 
the principal investigator, as is customary at the CFHT. We used these
preprocessed frames, and subtracted the sky from the near-infrared
images by averaging over the ten frames that had been taken most closely 
in time to a given frame, then subtracted the average from this science 
frame. Individual frames were aligned relative to each other and to the 
world coordinate system with the astrometric tools {\sc Swarp} and 
{\sc Scamp} \citep[][]{bertin10a,bertin10b}, resampled to 0.3\arcsec\ pixel 
scale, cropped to 5\arcmin$\times$ 5\arcmin, and flux calibrated relative 
to the Sloan Digital Sky Survey \citep[SDSS,][]{alam15} and the Two Micron 
All-Sky Survey \citep[2MASS,][]{skrutskie06} for the optical and near-infrared 
images, respectively. We obtained a relative calibration and zero-point 
uncertainties below 0.05~mag in each band, by fitting the spectral energy 
distribution of nearby non-saturated stars with blackbody spectral energy 
distributions. 

The IRAC images were obtained as part of program 10010 (PI: Nesvadba)
during \textit{Spitzer} observing cycle~10 on 8 July 2014, and were
observed through the 3.6~$\mu$m and 4.5~$\mu$m filters as part of the
warm mission. The total observing time per band was 1200~s, composed
of individual exposures of 100~s duration. Basic calibrated data were 
released after a preliminary processing conducted by the \textit{Spitzer}
Science Center standard pipeline. The dark currents and flat fields
were automatically calibrated and subtracted during this stage. In
both channels the frames were flux calibrated, combined into mosaics
with 0.60\arcsec ~pix$^{-1}$ sampling, corrected for cosmic ray artifacts 
and astrometrically calibrated with external 2MASS catalogs. After they
were released to us, we again used {\sc Swarp} and {\sc Scamp} to put
these images onto a common reference frame with our ground-based data.

\begin{figure*}
\centering
\includegraphics[width=0.8\textwidth]{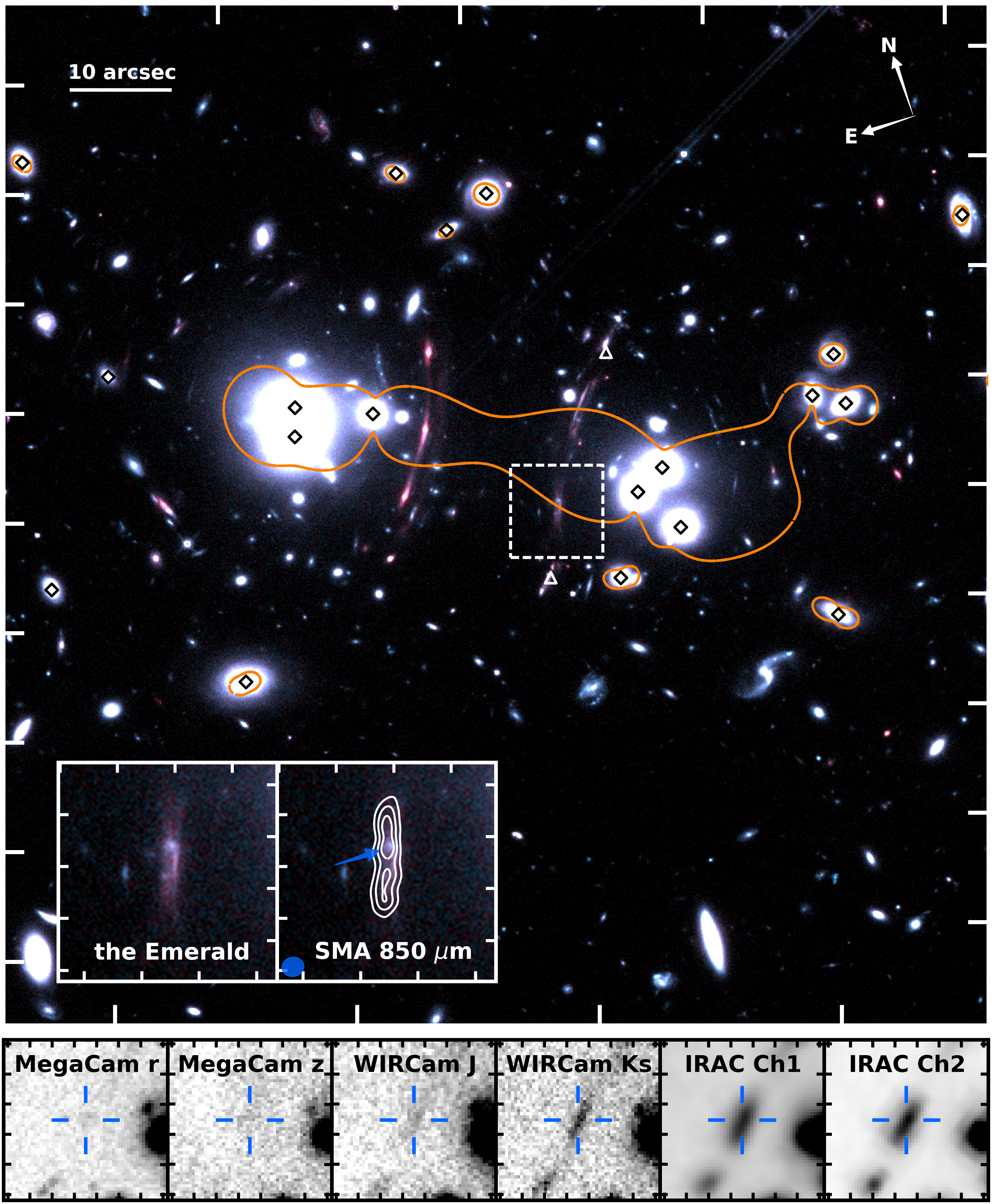}
\caption{{\it Top:} \textit{HST}/WFC3 F110W and F160W band two-color image of the
{\it Planck}'s Dusty GEMS PLCK\_G165.7+49.0 and surrounding field of view. 
The orange solid line shows the critical curve at $z=2.236$, for the 
gravitational lensing potential from our best-fitting lens model, derived using 
the position and multiplicity of the lensed images identified with \textit{HST} 
and the PdBI (see text and Table~\ref{tab:lensima}). Black diamonds indicate 
the members of the foreground structure that were included in the lens model,
and white triangles mark the position of the two compact submm emitters.
The left inset shows a 9\arcsec~$\times$~9\arcsec\ wide enlargement of the 
submm arc at $z=2.236$, ``the Emerald''. In the right inset, we show the dust 
continuum at 850~$\mu$m from the SMA, with contours starting at $4\sigma$ and 
increasing in steps of 4$\sigma$; the beam size is shown in the lower left 
corner and the blue arrow marks the position of the stellar continuum clump 
further discussed in the text. The bar in the upper left corner shows a 
projected distance of 10\arcsec, corresponding to 50~kpc at $z=0.348$ and 
84~kpc at $z=2.236$. {\it Bottom:} 15\arcsec~$\times$~15\arcsec\ postage 
stamps centered on the Emerald (blue symbols) from our optical and infrared 
imaging obtained with CFHT and \textit{Spitzer}.}
\label{fig:panorama} 
\end{figure*}

\subsection{Aperture photometry}
\label{ssec:photometry}

We selected our sources from the $K_{\rm s}$-band imaging to approximate a 
mass selection at intermediate redshifts, and used {\sc Sextractor} 
\citep[][]{bertin96} to measure aperture magnitudes of 3.9\arcsec\ diameter 
in all CFHT bands. We verified carefully that these apertures were large 
enough to minimize flux losses, while being small enough that sources were 
not blended. For extended sources, we used the corrected 
isophotal magnitudes from {\sc Sextractor} computed down to the 3$\sigma$ 
isophotes. Positional uncertainties relative to the $K_{\rm s}$ band have 
${\rm rms} \simeq 0.1\arcsec$, about 10--15\% of the FWHM size of the point 
spread function (PSF). 

The IRAC images have substantially larger PSFs, with FWHM sizes of around 
1.7\arcsec, so that blending becomes more important, in particular in the 
denser regions of our foreground structures. We measured magnitudes within 
the same apertures of 3.9\arcsec\ as for the ground-based data, and applied 
aperture correction factors of 1.4 and 1.5 in the 3.6 and 4.5~$\mu$m bands, 
respectively, following \citet{barmby08} and \citet{martinache18}. Positional 
uncertainties are between 0.2\arcsec\ and 0.5\arcsec\ relative to the 
$K_{\rm s}$-band image. 

The final catalog includes 737 objects down to 3$\sigma$ limiting AB 
magnitudes of 25.5, 23.7, 23.9 and 23.0~mag in the $r$, $z$, $J$, and $K_{\rm s}$ 
bands, respectively. In the 3.6~$\mu$m and 4.5~$\mu$m \textit{Spitzer} 
channels, the limiting magnitudes are 23.5 and 23.4~mag.

\subsection{SMA 850 $\mu$m interferometry}
\label{ssec:obs.sma}

The 850~$\mu$m continuum (345~GHz) of PLCK\_G165.7+49.0 was observed
with the SMA on 12 December 2013, with a total integration time of seven
hours as part of program 2013B-S050 (PI: Nesvadba) in the compact (COM)
configuration at 339.15~GHz, and with another seven hours as part of program 
2016B-S005 (PI: Nesvadba) in the extended (EXT) configuration at the 
same frequency on 25 March 2017. Combining data from both runs gives a 
beam with FWHM size of 0.9\arcsec~$\times$~0.75\arcsec\ at PA~=~39$^\circ$. 
The data were taken under excellent conditions, with precipitable water 
vapor below 2~mm, and with individual scan durations of 30~s~beam$^{-1}$. 
In the COM run, we used Callisto as flux calibrator, and \object{1153+495} 
and \object{1159+292} were used as phase calibrators with the bandpass 
calibrated on \object{3C~279}. In the EXT run, we used \object{1159+292} 
and \object{1146+399} as phase calibrators, Titan as flux calibrator, and 
\object{3C84} to calibrate the bandpass.

Data were reduced and calibrated with the {\tt MIR} package, and Fourier 
transformed and deconvolved with {\tt MIRIAD}. Images were created 
using ``Briggs'' weighting, with a parameter ${\rm robust=0.5}$ on the 
{\tt MIRIAD} task {\tt invert}, as is standard at the SMA. A comparison 
between the COM maps and the single-dish flux densities from SCUBA-2 
suggests that we recover at least 80--90\% of the total flux density at 
850~$\mu$m \citepalias[][]{canameras15}. We reached an rms noise of about 
$1.6$~mJy~beam$^{-1}$ using the SMA.

We also observed PLCK\_G165.7+49.0 with the very extended (VEXT) 
configuration of SMA on 20 January 2015, with a total integration time of 
seven hours and 0.3\arcsec\ beam size. The rms reached with these observations 
was 1.75~mJy beam$^{-1}$. PLCK\_G165.7+49.0 was not detected in these 
observations, suggesting that the flux was either resolved out or that 
the surface brightness was too faint even at the center of the clumps 
to be measured with the small beam size. At any rate, the non-detection
suggests that there are no high surface-brightness clumps of dust
emission in PLCK\_G165.7+49.0 that are more compact than about
0.3\arcsec\ in the image plane.

\subsection{IRAM CO(4--3) interferometry and spectral line fitting}
\label{ssec:obs.iram}

The CO(4--3) line from PLCK\_G165.7+49.0 was observed with the PdBI 
with six antennas in the B configuration in Band~2 on March 19 2014. 
The shortest and longest baselines in the
data set are 88~m and 452~m and thus, the observations are sensitive 
to scales smaller than approximately 5\arcsec. The phase center of 
the observations was located at $\alpha$=11h27m14.60s and
$\delta$=+42\degr28\arcmin25.0\arcsec. The 2~mm receivers were tuned 
to a sky frequency of 142.47~GHz, corresponding to the rest-frame 
wavelength of CO(4--3) at a redshift $z=2.23606$. The WideX correlator 
with its 3.6~GHz bandwidth at a spectral resolution of 1.95~MHz 
provided a velocity coverage of 7500~km~s$^{-1}$ with 4.1~km~s$^{-1}$ 
wide channels. The bright quasar \object{3C~84} was used for bandpass 
calibration, \object{LkHa~101} was observed as primary flux calibrator, 
and we regularly observed the nearby quasars \object{1150+497} and 
\object{1128+385} for gain and phase calibration.

The data were calibrated and imaged within {\tt GILDAS/CLIC} and 
{\tt MAPPING}\footnote{\url{http://www.iram.fr/IRAMFR/GILDAS}}. A few 
outliers in the visibilities of channel 42 were removed using the 
{\tt uv\_clip} task in {\tt MAPPING}. To image the data, we used 
the standard {\tt clean} procedure together with a mask that was 
carefully adapted to each individual frequency channel, and applied 
``natural'' weighting. This resulted in a beam size of 
0.76\arcsec~$\times$~0.75\arcsec\ at a position angle of 92\degr. The 
data reach an rms of 0.6~mJy beam$^{-1}$ per spectral channel, which 
are 42~km~s$^{-1}$ wide. After calibration and imaging the data cube 
was exported as a fits file for analysis.

We used {\tt MPFIT} \citep{markwardt09} to construct maps of the line 
fluxes, local velocities relative to $z=2.23606$, and FWHM line widths 
from our CO(4--3) data cube. These maps are shown in Fig.~\ref{fig:comaps}. 
In a first step, we fit a single Gaussian component to spectra extracted 
from apertures of 3$\times$3 spatial pixels, or 
0.6\arcsec~$\times$~0.6\arcsec, slightly less than the beam size. This 
maximizes the signal-to-noise ratio while causing no loss of spatial 
information. We only fit pixels in which the line was detected at 
$\ge 3\sigma$. In a small region near the critical line and extending to 
about 1.3\arcsec\ from it on either side (see Fig.~\ref{fig:comaps}), 
single-Gaussian fits lead to significant residuals ($\ge 3 \sigma$). We 
adopted a two-component fit in this region to include this component in 
our analysis. The integrated spectrum and maps corresponding to this
secondary component are shown in Fig.~\ref{fig:windspec} and in the 
lower panel of Fig.~\ref{fig:comaps}, respectively. The total CO flux
extracted from this region is $\mu~I_{\rm CO(4-3)}=19.5\pm 0.4$~Jy~km~s$^{-1}$,
corresponding to a total luminosity, $\mu~L^\prime_{\rm CO(4-3)}=2.9\times
10^{11}$~K~km~s$^{-1}$~pc$^2$, uncorrected for the gravitational 
magnification factor $\mu$.

\begin{figure*}
\centering
\includegraphics[width=0.6\textwidth]{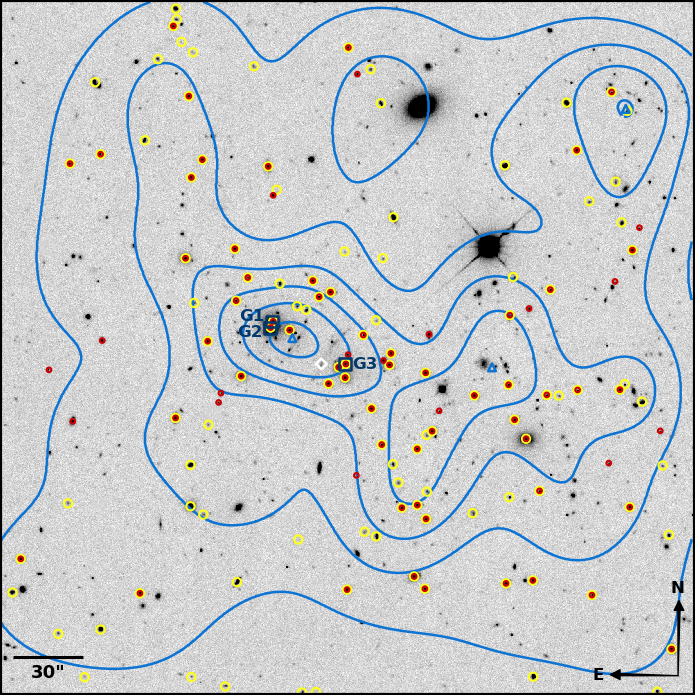}
\caption{Distribution of near-infrared sources toward PLCK\_G165.7+49.0, including 
candidate members of the foreground structure. The grayscale shows the CFHT 
$K_{\rm s}$-band image and blue contours indicate local overdensities from
the adaptive kernel density estimate, from 1 to 6$\sigma$. A main peak at 
6.4$\sigma$ near the Emerald and two secondary peaks at $>4\sigma$ are found 
(blue triangles), indicating a particularly rich environment. The position 
of the Emerald (the submm arc) is shown as a white diamond. Red and yellow 
circles indicate galaxies that fall onto the red sequence, or have photometric 
redshifts consistent at the 2$\sigma$ level with $z=0.35$, respectively. The 
three galaxies marked G1, G2 and G3 have spectroscopic redshifts around $z=0.35$ 
available in the SDSS, as reported in Table~\ref{tab:sdssspectra}. We show the 
5\arcmin~$\times$~5\arcmin\ field-of-view used to characterize the foreground 
mass distribution. North is up and east is to the left.}
\label{fig:akde}
\end{figure*}

\subsection{Ancillary data sets}
\label{ssec:obs.ancillary}

We obtained \textit{HST}/WFC3 imaging through the F110W and F160W filters 
for a subset of the GEMS through program 14223 (PI: Frye). For details of 
the observations and data reduction see \citet{frye18}. A major 
justification of that observing program was to refine the strong lensing 
model obtained with {\sc Lenstool} as presented in the present paper, which 
depends critically on measuring accurate positions, colors and morphologies 
of the multiply imaged sources from high-resolution imaging, as \textit{HST} 
ideally provides. We therefore included these data in our photometric 
catalogs, and based our lens model on the colors and positions measured 
from these \textit{HST} images, which have a point spread function with a 
FWHM size of about 0.15\arcsec.

We also used spectra from the 12th data release of the Sloan 
Digital Sky Survey \citep[][]{alam15}, available for four targets, to validate 
our photometric redshifts through spectroscopy, and to confirm the cluster 
redshift obtained through a red-sequence analysis. Three red-sequence galaxies 
also have SDSS spectra, including the two main galaxies of the groups directly 
adjacent to the gravitationally lensed arcs (see Fig.~\ref{fig:akde}). Their 
spectral properties are listed in Table~\ref{tab:sdssspectra}. 

\section{Multiple extended arcs behind a rich foreground environment}
\label{sec:foreground}

Since the GEMS were unknown targets previous to our follow-up observations, 
and do not fall into well characterized regions of the night sky, we must 
apply some care in characterizing the foreground mass distributions that are 
responsible for the gravitational lensing effect. 

Figure~\ref{fig:panorama} shows a wide view of the field around the line of 
sight toward PLCK\_G165.7+49.0, revealing multiple red, strongly gravitationally 
lensed arcs. These arcs are mainly distributed along two extended curves with 
projected lengths of 28.5\arcsec\ and 21\arcsec, respectively, which we will 
refer to as the western and eastern arcs, respectively, and which fall between 
two groups of early-type galaxies that have similar SDSS spectroscopic redshifts
around $z=0.35$ (Table~\ref{tab:sdssspectra}) and might represent 
substructure within a galaxy cluster. This makes the field of PLCK\_G165.7+49.0 
the richest environment toward any of our GEMS. Both arcs are also resolved in 
the narrow direction, with widths of typically about 0.6\arcsec. 

Only subcomponents of the western arc are associated with dust continuum and 
CO line emission, probed with the SMA and PdBI, and will be discussed in more
detail in Sect.~\ref{sec:lemeraude}. The stellar continuum emission along the 
western arc is globally diffuse and filamentary, with one pronounced 
clump near the center of the component aligned with the submillimeter 
emission (see Fig.~\ref{fig:panorama}). One subcomponent of the western arc 
is particularly bright in the submillimeter, producing most of the dust 
continuum emission of PLCK\_G165.7+49.0. Its integrated 850~$\mu$m flux 
density is 48~mJy, about two thirds of the total 850~$\mu$m flux density of 
71.6~mJy measured with the SMA within the half-power beam width of SCUBA-2. 
In the following discussion we refer to this submillimeter arc as 
``the Emerald''.

The CO interferometry shows that this arc probes a single source at a 
redshift $z=2.236$, with a velocity gradient consistent with the presence
of two merging images, as further discussed in Sect.~\ref{sec:lenstool}. 
Toward the southeast and northwest are two additional, compact and more 
moderately magnified CO emitters, which also fall within the 
20\arcsec--30\arcsec\ beam of SPIRE and typical single-dish observations 
in the submillimeter and millimeter wavelength range. We will refer to 
these sources as ``Co-S'' and ``Co-N'' (see Fig.~\ref{fig:panorama}). 
They are also detected in CO(4--3) and are at similar redshifts to the 
Emerald.

We followed \citet{song12a,song12b} in adopting a three-way approach 
to characterize this environment in a more rigorous, quantitative
way. Firstly, we calculated the projected density along the line of
sight, using the adaptive kernel density estimation (AKDE) algorithm,
which adopts a non-parametric, scale-independent smoothing technique
to calculate the local density around each detected source
\citep[see][]{ferdosi11,pisani96,valtchanov13}. We used the package
already described by \citet{valtchanov13} to determine the position
and significance of overdensity peaks. Secondly, we searched for the
presence of a red sequence in optical and near-infrared color-magnitude 
diagrams, which would be the clearest signature of a dense region in 
three dimensions, and provide robust and accurate redshift constraints
\citep[][]{fassbender08,fassbender11}. Thirdly, we used the publicly
available Bayesian photometric redshift package \citep[{\tt BPZ},][]{benitez00} 
to estimate photometric redshifts and study the redshift distribution 
along the line of sight. We will start by describing the analysis of the 
local source density projected onto the sky around the sightline toward
PLCK\_G165.7+49.0, and come back to the second and third steps in the
following two subsections.

\subsection{Local projected source density}
\label{ssec:akde}

As a first step to quantify the nature of this structure, we estimated
the density distribution of sources selected in the CFHT $K_{\rm s}$-band
within a field-of-view of 5\arcmin~$\times$~5\arcmin\ around
PLCK\_G165.7+49.0. We applied the AKDE on the 519 detections, finding 
a density peak of 6.4$\sigma$ significance at about 10\arcsec\ from 
the Emerald, and secondary peaks of 4.6$\sigma$ and 4.1$\sigma$ at 
separations of 1.3\arcmin and 2.9\arcmin, respectively. We took 
advantage of the exceptionally large field-of-view of WIRCAM of
20\arcmin~$\times$~20\arcmin\ to estimate the background standard
deviation of the AKDE, $\sigma$, at the same depth and observing
conditions. We also applied the AKDE to 5\arcmin~$\times$~5\arcmin\ 
wide areas randomly distributed over the WIRCAM field-of-view, but far 
away from the GEMS and surrounding galaxy overdensities, and adopted 
the median density in these fields as a conservative appoximation of 
the source density in the field. 
Comparison with the density contours obtained in WIRCAM's entire 
field-of-view indicated that the size of the field neither affects the 
position of the AKDE peaks nor the morphology of the overdensity contours, 
as expected for a scale-free method.

The density contours are shown in Fig.~\ref{fig:akde}. In the following 
subsections, we further characterize the most significant overdensity 
associated with PLCK\_G165.7+49.0. Its 4$\sigma$ contours define a region 
with a projected major axis length of 73\arcsec.

\subsection{Red sequence analysis and cluster redshift}
\label{ssec:redsequence}

Massive galaxy populations in dense environments fall within a tight,
well defined region in optical and near-infrared color-magnitude
diagrams. This ``red sequence'' is a signature of their early, rapid,
and very uniform formation history for a short epoch at high
redshift, followed by passive evolution for most of cosmic history. 
The tightness and uniformity of the red sequence makes it an excellent 
tool for determining redshifts of the overall structure, often with 
uncertainties better than $\Delta z = 0.05$ out to redshifts $z \simeq 1$ 
\citep[e.g.,][]{fassbender11}. In the case of the GEMS, we can use the red 
sequence to search for the presence of one or multiple massive structures 
along the line of sight, which could contribute to the gravitational 
magnification of the GEMS, and to help determine their redshifts.

\begin{figure}
\centering
\includegraphics[width=0.48\textwidth]{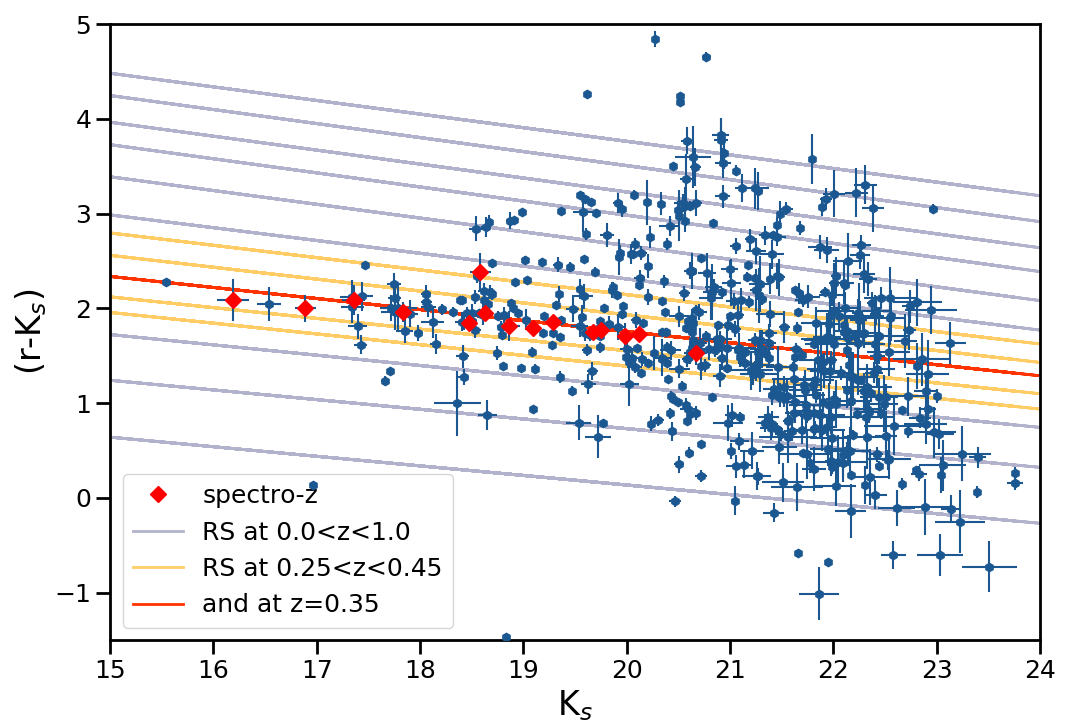}
\caption{Color-magnitude diagram of a 5\arcmin~$\times$~5\arcmin\ field-of-view 
surrounding PLCK\_G165.7+49.0, showing the $r-K_{\rm s}$ color versus 
$K_{\rm s}$-band magnitude. Gray and yellow lines show the expected position 
of the red sequence for a range of redshifts between $z=0$ and $z=1.0$ and 
between $z=0.25$ and $z=0.45$, respectively. The red line indicates the best-fit 
sequence at $z=0.35$. Red stars indicate galaxies which have spectroscopic 
redshifts from the SDSS or \citet{frye18} falling in the range 
$z=0.350 \pm 0.025$.}
\label{fig:cmdiagram}
\end{figure}

The $r-K_{\rm s}$ versus $K_{\rm s}$ color-magnitude diagram of the field-of-view
around PLCK\_G165.7+49.0 in Fig.~\ref{fig:cmdiagram} shows a clear red 
sequence, with colors $r-K_{\rm s} \simeq 2.0$, which is associated with the 
massive early-type galaxies seen in Fig.~\ref{fig:akde}, and which we can 
use to determine a redshift of the overall structure, to measure its extent, 
and to identify the member galaxies.

We followed \citet{song12a,song12b} in modeling synthetic spectra of
early-type galaxies using the stellar population synthesis tool of
\citet{bruzual03}, and to populate an artifical red sequence matched to
that observed in the Coma cluster at $z=0.0023$ \citep[][]{eisenhardt07}. 
More explicitly, we adopted short starbursts with e-folding time 
$\tau=50$~Myr and a Chabrier initial mass function, starting at $z=3$, 
and followed by passive evolution until today. We used six different 
templates from the Padova library covering a wide range of metallicities 
from 0.05 to 2.5 times the solar metallicity, and extracted the spectral 
energy distributions for several ages corresponding to redshifts $z=0.0$ 
to $z=1.0$ in steps of $\Delta z = 0.05$. These SEDs were then rescaled 
to best reproduce the color and slope of the red sequence measured on 
the Coma cluster, using the eight color-magnitude diagrams from
\citet{eisenhardt07} with the same set of filters.

We redshifted these SEDs to our grid of redshifts out to $z=1.0$, and
convolved them with the transmission curves of the CFHT filters to obtain 
a range of red sequence models in each color-magnitude diagram. For 
PLCK\_G165.7+49.0, which shows a pronounced red sequence at $r-K_{\rm s} 
\simeq 2$, we find a best-fitting model at $z=0.35$, in excellent agreement
with the spectroscopic redshifts given in the SDSS for the three galaxies 
within the groups adjacent to the submm components 
(Table~\ref{tab:sdssspectra}) and with those presented in \citet[][see 
Fig.~\ref{fig:cmdiagram}]{frye18}. The galaxies with follow-up spectroscopy 
in \citet{frye18} were selected to fall onto the red sequence shown in 
Fig.~\ref{fig:cmdiagram}. Their redshifts, which all fall into the range 
expected from a massive cluster, are thus a direct confirmation of our 
photometry.

We selected galaxies that fall within the limits in the color-magnitude 
diagram set by our models for $z=0.3$ and $z=0.4$, with an additional 
magnitude cut to only include galaxies with $K_{\rm s}<21$. To reduce the 
number of interlopers, we performed a similar selection in the $J-K_{\rm s}$ 
versus $K_{\rm s}$ color-magnitude diagram, which also exhibits a narrow red 
sequence at $J-K_{\rm s} \simeq 0.5$, and considered only galaxies that fulfill 
both criteria as robust members of the cluster red sequence. We find 76 
members within the 5\arcmin~$\times$~5\arcmin\ field. Many of these members 
fall along a diagonal axis from the northeast to west (as seen in 
Fig.~\ref{fig:akde}). About half of the red-sequence galaxies lie in a 
2.5\arcmin~$\times$~1.5\arcmin\ wide region within the 3$\sigma$ AKDE contours, 
further indicating that the overdensity of NIR sources is not due to chance 
alignment along the line-of-sight. This suggests that PLCK\_G165.7+49.0 falls 
behind a massive, extended galaxy cluster with adjacent filament, which has 
so far not been identified by other surveys.

\begin{figure}
\centering
\includegraphics[width=0.48\textwidth]{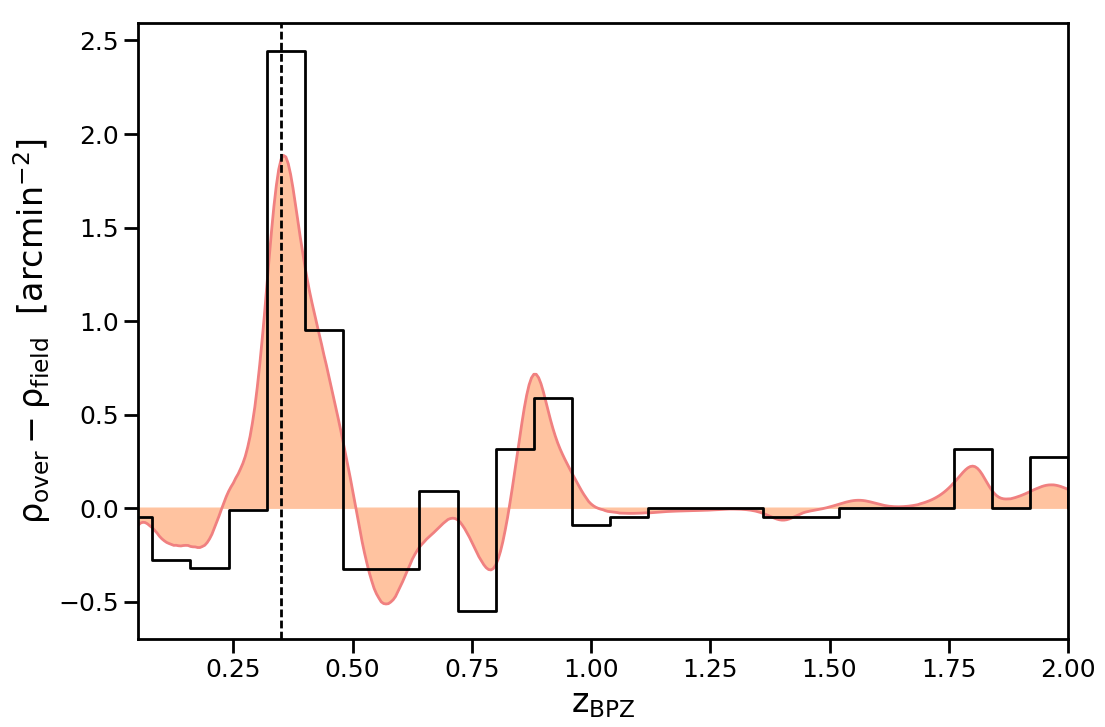}
\caption{Difference between the photometric redshift distribution of galaxies 
within 1\arcmin\ of PLCK\_G165.7+49.0, and the distribution obtained
in the rest of the 5\arcmin~$\times$~5\arcmin\ field, both normalized to 
the same area (black histogram). This shows a strong excess of sources in 
the inner field, which is consistent with $z \simeq 0.35$ (dotted vertical 
line). We used only the most robust photometric redshifts, and show the sum 
of their redshift probability distribution functions obtained with {\tt BPZ} 
(in orange).}
\label{fig:photoz}
\end{figure}

\subsection{Photometric redshifts}
\label{ssec:photoz}

We used the {\tt BPZ} package of \citet{benitez00} to estimate photometric 
redshifts from our $r$-, $z$-, $J$-, and $K_{\rm s}$-band photometry, which 
probes the 4000~\AA\ break for a redshift range of the member galaxies of the 
lensing structure of $z \simeq 0.2$--1.0, which appeared most likely prior to our 
analysis. We also included the shallower and bluer, publicly available SDSS 
$ugriz$ photometry to improve the robustness of our redshift estimates for the 
brightest targets, or upper limits in the bluer bands.

The {\tt BPZ} algorithm returns redshift probability distribution functions based 
on fitting a set of template SEDs. We used the standard set of templates provided 
by {\tt BPZ} and, since we are targeting fields with known bright FIR emitters, 
we also added two strongly reddened SEDs with $A_V=3$ and 5~mag, respectively. 
Including the IRAC photometry did not improve the robustness of our estimates, 
due to the greater photometric uncertainties, and because these wavebands probe 
relatively flat, featureless spectral regions of the SEDs for galaxies at low 
and intermediate redshifts, where most of our sources lie. Because of this, we 
did not include these two bands in our analysis.

In the 5\arcmin~$\times$~5\arcmin\ field surrounding PLCK\_G165.7+49.0, we 
identified 187 galaxies with reliable photometric redshifts (parameters ODDS 
$>0.9$ and $\chi^2<10$ in {\tt BPZ}). Comparison between these photometric 
redshifts and spectroscopic redshifts of 30 sources, taken from the SDSS and 
follow-up spectroscopy with MMT/HECTOSPEC and Gemini/GMOS \citep{frye18},
showed that our estimates are robust, with an average scatter 
$ \left| z_{\rm spec}-z_{\rm BPZ} \right|/(1 + z_{\rm spec}) \simeq 0.09$.

As shown in Fig.~\ref{fig:akde}, about 80\% of the red-sequence galaxies have 
photometric redshifts consistent at the 2$\sigma$ level with $z=0.35$, which 
makes them good candidates for being members of the foreground structure.
We computed the difference between the redshift distribution of sources
within 1\arcmin\ of PLCK\_G165.7+49.0 and associated with the AKDE peak, 
and the redshift distribution of sources in the rest of the field, after 
normalizing to the same area (Fig.~\ref{fig:photoz}). This shows a strong 
excess of sources at $z=0.3$--0.4 toward PLCK\_G165.7+49.0 and the presence 
of a massive structure in this redshift range, which is consistent with 
the redshift of $z=0.35$ found from the red sequence analysis and the 
spectroscopic redshifts of galaxies G1, G2 and G3 from SDSS.

\section{Gravitational lens modeling}
\label{sec:lenstool}

The {\it Planck}'s Dusty GEMS PLCK\_G165.7+49.0 falls behind a very rich, 
so far unexplored galaxy 
environment, as discussed in the previous section. The western and eastern 
extended arcs seen in the $K_{\rm s}$-band, including the long-wavelength 
emission from the Emerald, Co-S and Co-N, fall between two compact groups 
of early-type galaxies at a common redshift $z=0.35$, which likely probe 
the inner region of a massive galaxy cluster. The small redshift offset in 
the SDSS spectra between the three galaxies of $\Delta z = 0.01$ shows that 
both are part of a single bound structure. Our red sequence and photometric 
redshift analysis also shows that this structure extends from northeast 
to west over about 3\arcmin, perhaps representing a massive filament 
(Fig.~\ref{fig:akde}). The structure is not detected in the \textit{Planck} 
catalog of clusters identified with the Sunyaev-Zel'dovich effect 
\citep[][]{planck15sz}, but has a faint X-ray counterpart in the Rosat 
All-Sky Survey \citep[][]{voges00}. All of this suggests that PLCK\_G165.7+49.0
is magnified by a massive dark-matter halo underlying the cluster at $z=0.35$, 
in addition to several member galaxies of this cluster at similar redshifts
forming a bimodal mass distribution, consistent with our spectroscopic and 
photometric constraints.

We performed a strong lensing analysis with {\sc Lenstool} \citep{jullo07}, 
by modeling the mass distribution toward the Emerald and neighboring 
western and eastern arcs. {\sc Lenstool} is a publicly available Bayesian 
lens modeling package, which uses the number of arclets detected in the image 
plane, their association in multiply imaged systems of the same regions in 
the source plane, and their positions relative to the critical line, in order 
to derive a best-fitting lensing potential that is responsible for the 
gravitational amplification.

\subsection{Identification of multiply imaged systems}

\begin{table*}
\centering
\begin{tabular}{lccccccccc}
\hline
\hline
Model & rms$_{\rm img}$ & Component & $\Delta$RA & $\Delta$Dec & $\epsilon$ & $\theta$ & $r_{\rm core}$ & $r_{\rm cut}$ & $\sigma$ \\
      & [\arcsec] & & [\arcsec] & [\arcsec] & & [deg] & [kpc] & [kpc] & [km s$^{-1}$] \\ [2mm]
\hline
Best & 0.21 & Large scale & 39.9 $\pm$ 4.0 & --20.3 $\pm$ 2.2 & 0.69 $\pm$ 0.07 & --33 $\pm$ 1 & 142 $\pm$ 10 & [500] & 1081 $\pm$ 112 \\
 & & G4 & [7.0] & [--1.5] & [0.1] & [--43] & [0.25] & [70] & 233 $\pm$ 14 \\
 & & $L^*$ galaxy & \dots & \dots & \dots & \dots & [0.25] & 59 $\pm$ 17 & 267 $\pm$ 17 \\ [2mm] 
\hline
Fixed center & 0.64 & Large scale & [31.2] & [--18.8] & 0.73 $\pm$ 0.03 & --31 $\pm$ 1 & 59 $\pm$ 5 & [500] & 759 $\pm$ 25 \\
 & & G4 & [7.0] & [--1.5] & [0.1] & [--43] & [0.25] & [70] & 293 $\pm$ 9 \\
 & & $L^*$ galaxy & \dots & \dots & \dots & \dots & [0.25] & 40 $\pm$ 19 & 213 $\pm$ 30 \\ [2mm]
Faint arcs & 0.43 & Large scale & 28.6 $\pm$ 1.0 & --14.4 $\pm$ 0.7 & 0.70 $\pm$ 0.09 & 32 $\pm$ 1 & 82 $\pm$ 12 & [500] & 851 $\pm$ 58 \\
 & & G4 & [7.0] & [--1.5] & [0.1] & [--43] & [0.25] & [70] & 243 $\pm$ 9 \\
 & & $L^*$ galaxy & \dots & \dots & \dots & \dots & [0.25] & 63 $\pm$ 23 & 296 $\pm$ 28 \\ [2mm]
Non-cored & 0.34 & Large scale & 24.9 $\pm$ 0.5 & --11.6 $\pm$ 0.5 & 0.64 $\pm$ 0.09 & 31 $\pm$ 2 & [20] & [500] & 573 $\pm$ 26 \\
 & & G4 & [7.0] & [--1.5] & [0.1] & [--43] & [0.25] & [70] & 284 $\pm$ 11 \\
 & & $L^*$ galaxy & \dots & \dots & \dots & \dots & [0.25] & 133 $\pm$ 29 & 301 $\pm$ 16 \\ [2mm] 
NFW & 0.30 & Large scale & 37.8 $\pm$ 1.6 & --20.1 $\pm$ 0.9 & 0.68 $\pm$ 0.05 & --32.5 $\pm$ 0.6 & 3.4 $\pm$ 0.7 $^{(1)}$ & 377 $\pm$ 53 $^{(2)}$ & \dots \\ 
 & & G4 & [7.0] & [--1.5] & [0.1] & [--43] & [0.25] & [70] & 253 $\pm$ 9 \\
 & & $L^*$ galaxy & \dots & \dots & \dots & \dots & [0.25] & 74 $\pm$ 21 & 296 $\pm$ 24 \\ [1mm]
\hline
\end{tabular}
\caption{Parameters of the foreground mass distribution inferred by {\sc Lenstool}, 
for our best lensing model and the four alternative models. ``Large scale'' 
refers to the extended dark-matter halo associated with the lensing 
structure. Positional offsets are given in arcseconds relative to 
$\alpha$=11h27m16.6s and $\delta$=+42\degr28\arcmin38.8\arcsec. The 
ellipticity of the mass distribution, $\epsilon$, is given as 
$(a^2-b^2)/(a^2+b^2)$. Parameters in brackets are fixed, and errors 
correspond to 1$\sigma$ confidence intervals for the best-fit parameters. 
For the NFW model, (1) and (2) refer to the concentration parameter, $c$, 
and the scale radius, $r_{\rm s}$, respectively.}
\label{tab:lensopt}
\end{table*}

The {\sc Lenstool} software relies on two sets of observational parameters, 
which can be obtained from the imaging -- the number and position of multiple 
lensed images, and the position, brightness, and structural parameters 
of the foreground lensing galaxies. The parametrization of the dark-matter 
halo underlying the galaxy cluster must also be provided. 

In the following analysis, we use the \textit{HST}/WFC3 F110W and 
F160W images also presented by \citet{frye18} to measure the position 
of the faint lensed arcs and foreground galaxies, and our IRAM CO(4--3) 
interferometry to constrain the positions of the images for the 
long-wavelength emitters. All images are magnified by the same lens
regardless of the wavelength in which they are studied, so combining
both sets of constraints in a joint analysis provides the most robust
model of the underlying mass distribution. {\sc Lenstool} determines
the best-fit model by minimizing the positional offsets between the
measured and reconstructed image positions. We have considered a model
adequate when the rms of all offsets, rms$_{\rm img}$, is of order
0.1\arcsec--0.2\arcsec, the PSF of the \textit{HST}/WFC3 imaging.

The eastern and western extended arcs seen in the $K_{\rm s}$-band are 
resolved into several fainter arclets in the \textit{HST} imaging (see 
Fig.~\ref{fig:panorama} and \ref{fig:akde}). We identified compact clumps 
within each arclet and use their F110W--F160W color, their morphology, and 
spatial distribution to combine them into seven multiply imaged systems 
with unambiguous associations. Given the lack of spectroscopic redshifts 
for these near-infrared-selected systems, we only used those with the most 
secure image configuration and colors. As shown in Fig.~\ref{fig:lensmodel},
the eastern arc comprises systems \#5, \#6 and \#7, with another nearby 
system \#8, and we identify systems \#2, \#3, and \#4 in the western arc.
The {\sc Lenstool} modeling constrains their redshifts, and we require
that $z<4.5$, as suggested by the non-detection of the Ly$\alpha$ emission 
line in Gemini/GMOS spectra \citep{frye18}. The similar colors and low 
angular separation between systems \#2, \#3 and \#4 suggest that they are 
subcomponents of a single background galaxy and we therefore assume 
a common redshift. 

In addition, we used the gas kinematics in PLCK\_G165.7+49.0 from the 
CO(4--3) emission line to probe the lensing configuration of the submm 
components. We identified the NIR counterpart of the Emerald in the F160W 
image with system \#1 at $z=2.236$. Figure~\ref{fig:comaps} shows that 
the direction of the velocity gradient is flipped in the northern and
southern parts of the submm arc, and so are the distributions of
the line widths shown in the same figure. This parity inversion is a
clear signature that the Emerald contains two merging images of the same
source. This is further shown by the small line widths in the center
of the arc, which would be difficult to explain for two independent,
and partially overlapping sources. Moreover, the line profiles and
velocity offsets of the two fainter, smaller submillimeter images (north 
and south from the arc, which we label Co-N and Co-S), have different
line profiles and velocity offsets (see Table~\ref{tab:othergalaxies} and
Fig.~\ref{fig:co10profiles}). While the overall properties of Co-N suggest
that this is another, less strongly magnified image of the same galaxy
that is also seen in the submm arc, Co-S seems to be another galaxy at 
18~kpc projected distance in the source plane. The properties of these 
galaxies are discussed in more detail in Sect.~\ref{sec:othergalaxies}.

In total, the arc and the other seven systems at unknown redshifts provide 
us with 28 constraints, which we can use to infer the properties of the 
lensing potential. The positions of all images included in the analysis are 
listed in Table~\ref{tab:lensima} and plotted in Fig.~\ref{fig:lensmodel}.

\subsection{Foreground mass distribution modeling}
\label{ssec:foregroundmassdistribution}

\begin{figure*}
\centering
\includegraphics[width=0.49\textwidth]{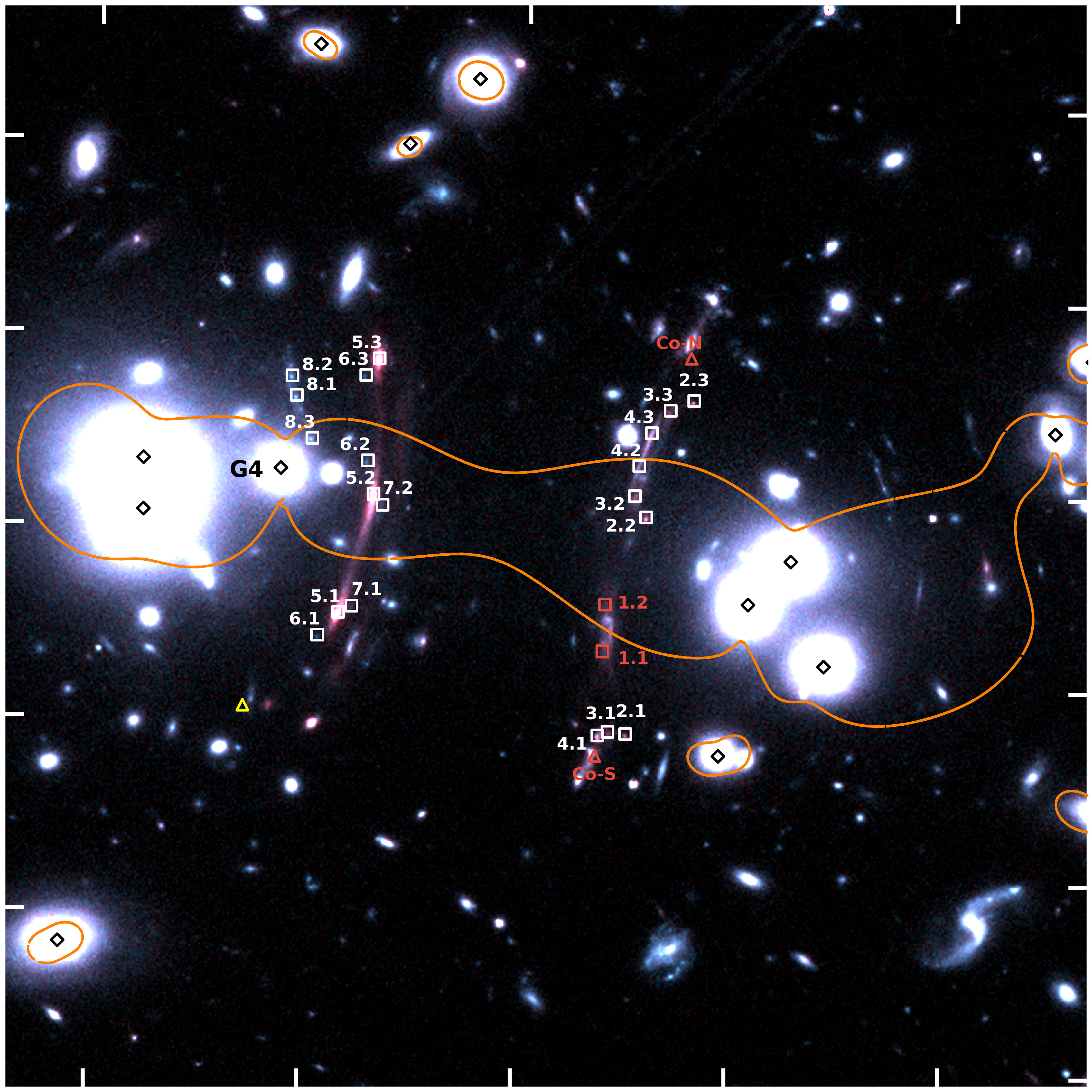}
\includegraphics[width=0.49\textwidth]{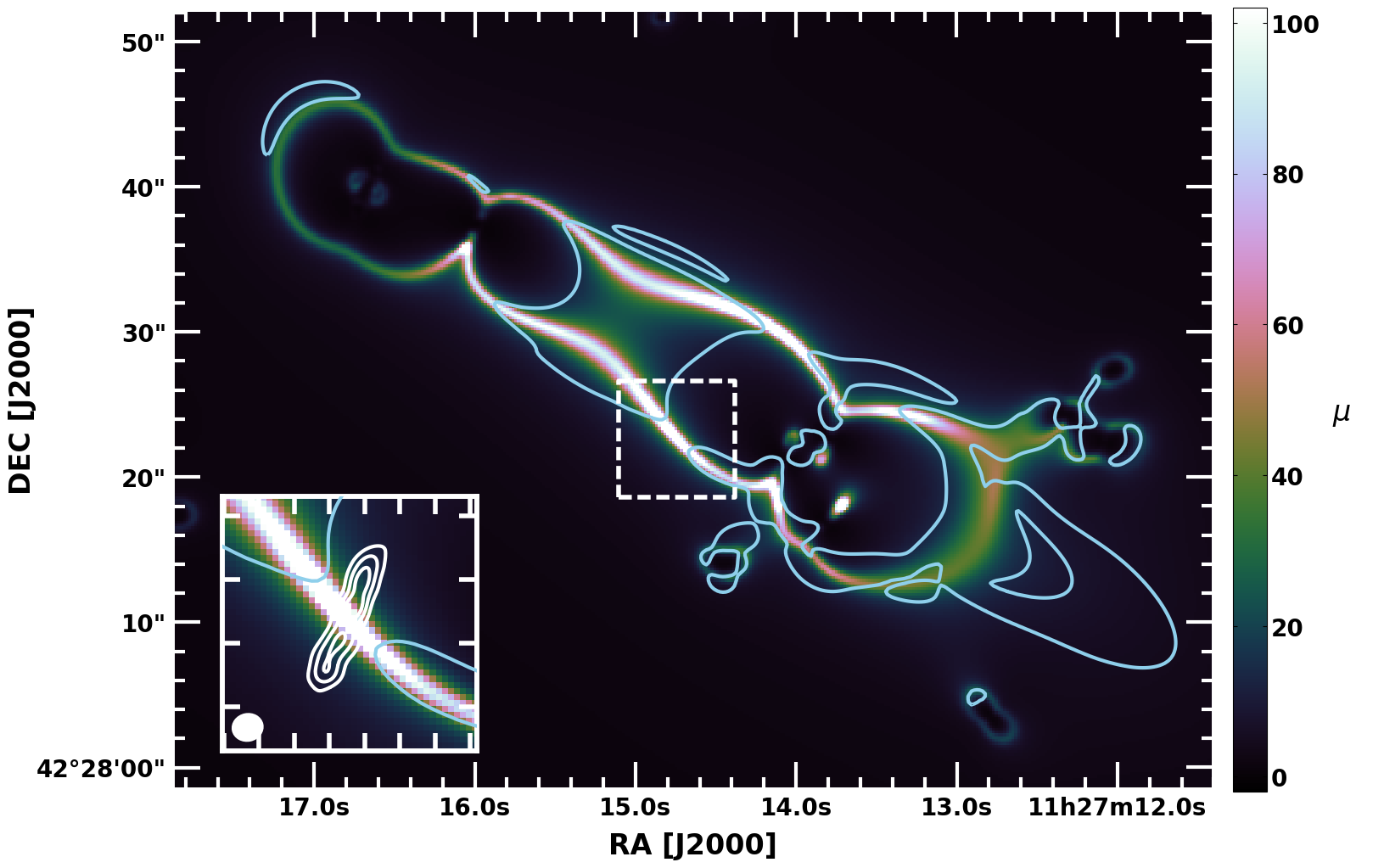}
\caption{{\it Left:} As Figure~\ref{fig:panorama}, with labels showing 
the multiple images identified for our gravitational lens modeling, 
either using IRAM CO(4--3) interferometry (red squares), or {\it HST} 
imaging (white squares). The yellow triangle indicates the position of 
the additional counter-image of system \#8 predicted by the best-fit 
model. {\it Right:} Magnification map obtained for the best-fitting 
lens model, by taking the median of the maps generated by 
{\sc Lenstool} for each MCMC realization. The inset shows the 
position of the SMA continuum emission from the Emerald (white 
contours, starting at $+4\sigma$ and increasing in steps of $+4\sigma$). 
Cyan contours indicate areas where the average of the relative 
difference in magnification between the best-fit and each of the 
four alternative models is higher than 30\%. The systematic errors 
induced by the different mass distributions in each model are much 
lower at the position of the Emerald thanks to the spectroscopic 
identification of the submm images.}
\label{fig:lensmodel}
\end{figure*}

The {\sc Lenstool} software inverts the lensing equation for the above input 
parameters, and derives the optimal set of parameters with Markov chain Monte 
Carlo (MCMC) simulations. We used 3000 MCMC iterations, and a large-scale
dark-matter halo underlying the overall structure as the main deflector,
together with the perturbations from smaller halos associated with the 
cluster members. We describe each mass component with a pseudo-isothermal 
elliptical mass distribution \citep[PIEMD,][]{eliasdottir07}, which has a 
radial profile characterized by a central mass surface density, a central
velocity dispersion, and core and cutoff radii, $r_{\rm core}$ and $r_{\rm cut}$ 
\citep{jullo07}. We note that the velocity dispersion of the PIEMD is not 
numerically identical to the physical velocity dispersion as measured
from stellar absorption lines \citep[see][for further details]{limousin07a}.

The large number of available constraints allows us to find the best
solutions for the projected position on the sky, the ellipticity and
the position angle of the main underlying dark-matter halo, as well as
its velocity dispersion and the core and cutoff radii. Dark-matter
halos are also assigned to individual galaxies on the red sequence, 
considering only galaxies within 1\arcmin\ of PLCK\_G165.7+49.0,
and rejecting those with photometric redshifts that are inconsistent 
with $z=0.35$ at $\ge 2\sigma$. The selection includes four galaxies 
with spectroscopic confirmation, three from the SDSS (G1, G2 and G3) 
and one from MMT/HECTOSPEC \citep{frye18}. In total, we include 21 
galaxies as perturbers in our model (shown with yellow and red circles 
in Fig.~\ref{fig:akde}).

To find the best-fitting foreground mass distribution we favor the simplest
parametrization that provides a rms consistent with our requirements.
We were unable to reproduce the position of lensed images and arcs 
without including a massive dark-matter component centered near the 
western group that underlies the overall structure. Given the lack of
spectroscopic redshifts for systems \#5, \#6, \#7 and \#8, we did not 
find a solution where adding a second potential associated with the 
eastern group would have improved the fit. Models using this second 
large-scale dark-matter halo are poorly constrained and result in mass 
components with unrealistically high ellipticities. We therefore describe 
the mass distribution toward PLCK\_G165.7+49.0 using a single large-scale 
dark-matter halo and galaxy-scale perturbers.

The cutoff radius of the main halo is not well constrained by the
lensing configuration, and we therefore set it to 500~kpc. We tested
carefully that the outcome of the lensing model depends only very 
weakly on the precise value of this parameter. We varied the ellipticity 
of the mass distribution between 0 and 0.8, the core radius between 30 
and 200~kpc, and the velocity dispersion between 400 and 2000~km~s$^{-1}$. 
We also allowed the position of the potential to vary within $\pm$10\arcsec\ 
with respect to the center of the western group. 

The position, ellipticity and position angle of individual galaxy halos 
are matched to the light profiles in the $K_{\rm s}$-band, while we let 
their cutoff radii and velocity dispersions scale with the galaxy 
luminosity following the two relationships
\begin{equation}
r_{\rm cut} = r^*_{\rm cut} \left( \frac{L}{L^*} \right)^{1/2} \\ {\rm and} \\
\sigma = \sigma^* \left( \frac{L}{L^*} \right)^{1/4}
\end{equation} \vspace{1mm}

\noindent Here $L$ is the luminosity of individual galaxies, and $L^*$ 
the characteristic luminosity of a galaxy at $z=0.35$. We adopted $K=16.0$ 
for an $L^*$ galaxy at $z=0.35$ \citep{depropris99} and varied the associated
characteristic cutoff radius, $r^*_{\rm cut}$, and velocity dispersion,
$\sigma^*$, between 50 and 150~kpc and between 150 and 300~km~s$^{-1}$,
respectively, following, for example, \citet{limousin07a} and 
\citet{richard14}. We held their core radius fixed at $r_{\rm core}=0.25$~kpc, 
as usually done in comparable studies in the literature 
\citep[e.g.,][]{brainerd96,limousin07b,richard14}. The rms of systems 
\#5 to \#8 is dominated by the dark-matter halo mass profile of a single 
foreground galaxy, labeled ``G4'' in Fig.~\ref{fig:lensmodel}. Accordingly, 
we determined the velocity dispersion of this halo separately. 

\begin{table*}
\centering
\begin{tabular}{lcccccccc}
\hline
\hline
Image ID & RA & Dec & $z_{\rm spec}$ & \multicolumn{5}{c}{$z_{\rm opt}$} \\
 & & & & Best & Fixed center & Faint arcs & Non-cored & NFW \\
\hline
 1.1 & 171.81167 & 42.472683 & 2.236 & -- & -- & -- & -- & -- \\
 1.2 & 171.81128 & 42.473342 & 2.236 & -- & -- & -- & -- & -- \\
 2.1 & 171.81197 & 42.471372 & -- & 2.2 $\pm$ 0.1 & 3.4 $\pm$ 0.3 $^{(1)}$ & 2.2 $\pm$ 0.2 $^{(2)}$ & 1.1 $\pm$ 0.5 & 1.9 $\pm$ 0.2 \\
 2.2 & 171.80998 & 42.474375 & -- & \arcsec & \arcsec & \arcsec & \arcsec & \arcsec \\
 2.3 & 171.80819 & 42.475792 & -- & \arcsec & \arcsec & \arcsec & \arcsec & \arcsec \\
 3.1 & 171.81230 & 42.471500 & -- & 2.3 $\pm$ 0.1 & 3.4 $\pm$ 0.3 $^{(1)}$ & 2.2 $\pm$ 0.2 $^{(2)}$ & 1.0 $\pm$ 0.4 & 2.0 $\pm$ 0.2 \\
 3.2 & 171.81004 & 42.474744 & -- & \arcsec & \arcsec & \arcsec & \arcsec & \arcsec \\
 3.3 & 171.80872 & 42.475778 & -- & \arcsec & \arcsec & \arcsec & \arcsec & \arcsec \\
 4.1 & 171.81252 & 42.471500 & -- & 2.4 $\pm$ 0.1 & 3.4 $\pm$ 0.3 $^{(1)}$ & 2.2 $\pm$ 0.2 $^{(2)}$ & 1.0 $\pm$ 0.4 & 2.0 $\pm$ 0.2 \\
 4.2 & 171.80974 & 42.475153 & -- & \arcsec & \arcsec & \arcsec & \arcsec & \arcsec \\
 4.3 & 171.80925 & 42.475558 & -- & \arcsec & \arcsec & \arcsec & \arcsec & \arcsec \\
 5.1 & 171.81668 & 42.474678 & -- & 2.2 $\pm$ 0.4 & 3.6 $\pm$ 0.4 & 2.0 $\pm$ 0.3 & 1.2 $\pm$ 0.6 & 1.8 $\pm$ 0.4 \\
 5.2 & 171.81513 & 42.476181 & -- & \arcsec & \arcsec & \arcsec & \arcsec & \arcsec \\
 5.3 & 171.81402 & 42.478100 & -- & \arcsec & \arcsec & \arcsec & \arcsec & \arcsec \\
 6.1 & 171.81725 & 42.474458 & -- & 2.1 $\pm$ 0.3 & 3.5 $\pm$ 0.3 & 2.0 $\pm$ 0.2 & 1.2 $\pm$ 0.5 & 1.7 $\pm$ 0.4 \\
 6.2 & 171.81499 & 42.476697 & -- & \arcsec & \arcsec & \arcsec & \arcsec & \arcsec \\
 6.3 & 171.81440 & 42.477931 & -- & \arcsec & \arcsec & \arcsec & \arcsec & \arcsec \\
 7.1 & 171.81637 & 42.474692 & -- & 2.3 $\pm$ 0.3 & 3.0 $\pm$ 0.3 & 2.0 $\pm$ 0.3 & 1.2 $\pm$ 0.5 & 1.8 $\pm$ 0.5 \\
 7.2 & 171.81503 & 42.475978 & -- & \arcsec & \arcsec & \arcsec & \arcsec & \arcsec \\
 8.1 & 171.81590 & 42.478017 & -- & 3.5 $\pm$ 0.3 & 1.0 $\pm$ 0.6 & 3.5 $\pm$ 0.4 & 1.8 $\pm$ 0.5 & 3.5 $\pm$ 0.3 \\
 8.2 & 171.81584 & 42.478325 & -- & \arcsec & \arcsec & \arcsec & \arcsec & \arcsec \\
 8.3 & 171.81591 & 42.477315 & -- & \arcsec & \arcsec & \arcsec & \arcsec & \arcsec \\
 9.1 & 171.82636 & 42.481387 & -- & $\oslash$ & $\oslash$ & 1.5 $\pm$ 0.8 & $\oslash$ & $\oslash$ \\
 9.2 & 171.82567 & 42.481899 & -- & $\oslash$ & $\oslash$ & \arcsec & $\oslash$ & $\oslash$ \\
\hline
\end{tabular}
\caption{Multiply imaged systems used to calculate the best-fit model of the foreground 
mass distribution. Colors and positions of most images are measured from the 
\textit{HST} imaging, expect for system \#1 which represents the PdBI CO(4--3) 
morphology of the Emerald, the bright submm arc. Systems \#1 to \#8 are the 
most securely identified from their position, color and morphologies and were 
used in each of the five models, while system \#9 was only considered in one 
alternative model and ignored in the remaining four ($\oslash$ symbols). Here 
$z_{\rm opt}$ refers to the redshifts derived from our lens models, and their 
1$\sigma$ confidence intervals, with indices (i) indicating systems that were 
fitted with a common redshift. Measuring spectroscopic redshifts for other 
multiply imaged systems would allow us to discriminate between models and to 
better constrain the foreground dark-matter distribution.}
\label{tab:lensima}
\end{table*}

The best-fit mass model reproduces the image positions with an rms$_{\rm img}$
of 0.21\arcsec. The modeled parameters of the PIEMDs are summarized
in Table~\ref{tab:lensopt}. The main dark-matter halo broadly follows
the light distribution, as it is oriented on the same axis as the
filamentary structure. Its ellipticity is close to the upper range expected 
from cosmological simulations \citep{despali17}. The main halo has a large 
core, with $r_{\rm core}>100$~kpc, and is offset by about 8\arcsec\ toward
the east from the center of the western group of galaxies. This is most likely 
due to the lack of constraints on the opposite side of the potential, as 
previously encountered in other studies of strong lensing clusters. The best-fit 
potential therefore corresponds to a bimodal mass distribution induced by 
the two groups of cluster members, with the large-scale dark-matter halo 
producing an additional convergence term. 

\begin{figure*} 
\centering
\includegraphics[width=1.00\textwidth]{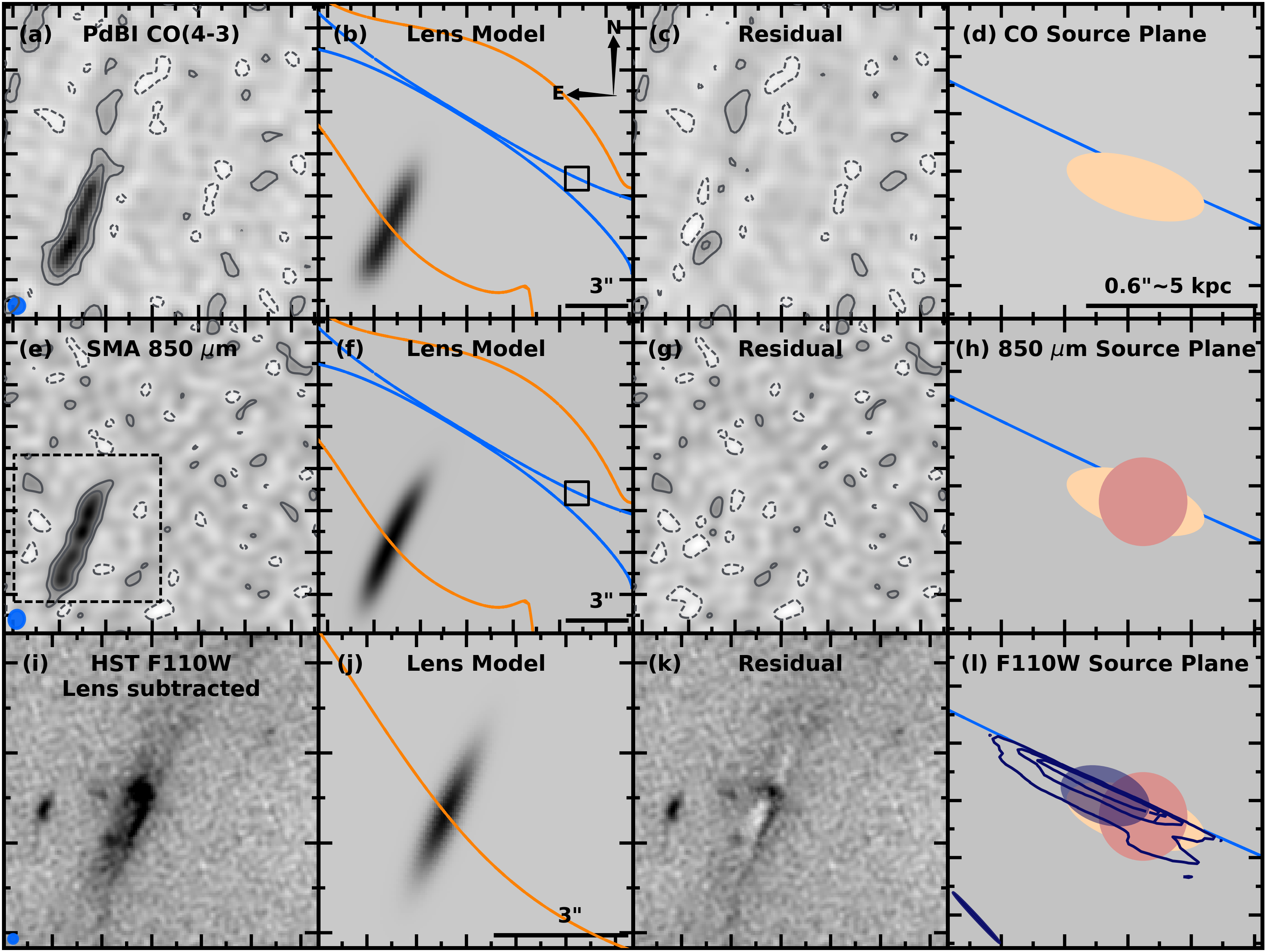}
\caption{Source-plane reconstruction of the gas, dust and stellar emission from the 
Emerald, using the best-fitting lens model of the massive foreground cluster. 
{\it Top row:} (a) PdBI map summed over the frequency channels of the 
CO(4--3) line. The black lines show contours at $-2\sigma$ (dashed lines), 
$+2$ and $+5\sigma$ (solid lines). The beam of the PdBI at the redshifted 
frequency of CO(4--3) is plotted in the lower-left corner. (b) Best-fit 
source model ray-traced to the image plane and convolved with the PdBI 
beam (see details in the text). Orange and blue solid lines show the 
critical and caustic curves at $z=2.236$, respectively, obtained from the 
best-fitting lens model. The box indicates the size and position of the 
enlarged region in the rightmost panel. (c) CO(4--3) residual and contours 
at $-2$, $+2$ and $+5\sigma$. (d) Best-fit model of the CO source, truncated 
at its FWHM (yellow ellipse). The bar in the lower-right corner illustrates 
the physical scales at $z=2.236$. {\it Center row:} (e) SMA 850~$\mu$m 
image in the extended configuration and $-2$, $+2$ and $+5\sigma$ contours 
(black lines). The SMA beam is shown in the lower-left corner. (f) Best-fit 
model convolved with the SMA beam in the image plane. (g) 850~$\mu$m residual 
and contours at $-2$, $+2$ and $+5\sigma$. (h) Best-fit azimuthally-averaged 
model of the 850~$\mu$m source, truncated at its FWHM (red circle). {\it Bottom 
row:} (i,j,k) Same as before, for the {\it HST} F110W imaging, within the 
field-of-view indicated in panel (e). (l) Best-fit model of the rest-frame 
optical stellar continuum in the source plane (blue ellipse), compared with 
those of CO(4--3) and 850~$\mu$m continuum. We also show the direct ray-tracing 
of the WFC3 F110W emission to the source plane (dark blue contours drawn at 
$+2$, $+3$ and $+4\sigma$), and the highly elongated PSF of the WFC3 image in 
the source plane (dark blue ellipse in the lower-left corner).}
\label{fig:sourceplane}
\end{figure*}

In Figure~\ref{fig:sourceplane} we show the critical line and image-plane 
morphology of the best-fit model reconstructed by {\sc Lenstool}, as 
well as the internal and external caustic lines and morphology of the 
Emerald in the source plane. In addition to images previously identified 
to constrain the mass distribution, the model predicts a fourth 
counter-image of system \#8, within 1\arcsec\ from a faint near-infrared 
source detected with \textit{HST} (see Fig.~\ref{fig:lensmodel}). It also 
predicts a third counter-image to the Emerald (system \#1), at a position 
consistent with that of Co-N in the dust and CO(4--3) maps. 

We explored the systematic errors on the magnification factors and source 
plane properties of the Emerald induced by the mass parametrisation and 
identification of multiple images, through deriving a grid of alternative 
models. We successively included and excluded some multiple image systems 
without spectroscopic redshifts, and some galaxies from the scaling relations. 
We also fitted and restricted different parameters of the large-scale PIEMD, 
and we tested different mass profiles \citep[following, e.g.,][]{limousin16}. 
After excluding models resulting in unphysical mass distributions and/or 
predicting a greater number of bright images than observed in \textit{HST} 
bands, we obtained four reasonable models with rms$_{\rm img}$ in the range 
0.2\arcsec--0.6\arcsec, with the following features:
\begin{enumerate}
\item we used a large-scale PIEMD with position fixed to the center of 
the western group;
\item we tested the effect of adding candidate counter-images on the 
opposite side of the foreground groups (see Table~\ref{tab:lensima});
\item we derived a non-cored model with $r_{\rm core}$ fixed to 20~kpc for 
the main halo;
\item we determined the impact of the mass-density slope degeneracy by 
replacing the main PIEMD with an NFW profile (while still describing 
individual member galaxies with PIEMDs).
\end{enumerate}

All models produce similar numbers of images at the correct positions, 
although they have different underlying mass distributions
(Table~\ref{tab:lensopt}) and predicted redshifts for the multiple
images (Table~\ref{tab:lensima}). Our current photometric identification 
of multiply imaged system thus induces some degeneracies in the model which 
prevent us from deriving robust constraints on the underlying dark-matter
distribution. However, this is not the main focus of this paper. Instead, we
estimated the corresponding systematic uncertainties in magnification, by 
computing the absolute value of the difference between the magnification 
map of the best-fit model and the four alternatives. Although the differences 
in the predicted magnification factors are significant for some systems of 
multiple images in Table~\ref{tab:lensima}, due to the different properties 
of the main lensing potential \citep[see also][]{limousin16}, they are not a 
major concern for the resulting gravitational magnification of submm
emitters, which is our focus in this paper. Figure~\ref{fig:lensmodel} 
illustrates that the average difference in magnification per pixel remains 
below 30\% toward all components of PLCK\_G165.7+49.0 including the Emerald;
this is because their spectroscopic identification in the submm forces all 
models to converge locally.

The F110W and F160W WFC3 images presented in Fig.~\ref{fig:panorama} 
illustrate that near-infrared emission from the Emerald is extended and 
diffuse, with fairly uniform surface brightness that varies by not more than 
about a factor of two or three, and over small scales comparable to the size of 
the PSF. The only exception is a brighter and very circular clump near the 
critical line, which has a significantly bluer color than the surrounding 
stellar emission (blue arrow on Fig.~\ref{fig:panorama}). The FWHM size of 
this clump is 0.45\arcsec~$\times$~0.33\arcsec\ and it is therefore well 
resolved with WFC3 at 0.15\arcsec~$\times$~0.15\arcsec\ PSF. While the 
diffuse continuum is clearly part of the high-redshift galaxy, the clump 
could in principle also be an intervening dwarf galaxy, perhaps a member 
of the main lensing cluster. The hypothesis of an intervening galaxy 
would in particular explain its very round, symmetric morphology. To 
evaluate the potential impact of this source for the lensing geometry, 
we ran {\sc Lenstool} for both hypotheses, finding that this clump, which 
might be a dwarf galaxy at $z=0.35$, would introduce a brightness 
difference of a factor of 0.8 between images \#1.1 and \#1.2 from the 
arc (as observed), but also distort image \#1.2 in ways that are not 
compatible with the stellar morphology seen with \textit{HST}. The 
impact of an intervening source at most other redshifts would be even 
smaller. While the observational evidence is thus not conclusive, neither 
scenario introduces major systematic uncertainties into our analysis.

\subsection{Magnification factors}
\label{ssec:magnificationfactors}

To obtain the magnification factors corresponding to our best-fit
model in each pixel in the image plane, we calculated a magnification
map for each MCMC realization at the redshift of the source, $z=2.236$, 
and computed the median of these maps. To determine the intrinsic 
properties of the stellar continuum, dust continuum at 850~$\mu$m, and 
molecular gas in the Emerald, and to account for possible differential 
magnification effects, we computed the luminosity-weighted magnification 
factors, $\mu$, for each component separately, using the same pixel scale 
as in each image. The three factors are deduced from the best-fitting 
lensing model combining gas and stellar constraints, by using pixels 
above a 3$\sigma$ threshold in the dust continuum and CO(4--3) line flux 
maps, and by using pixels included in the SExtractor segmentation map 
for the F160W band.

{\sc Lenstool} also computes distributions for the magnification
factors at the position of each image. For the submm arc, the
resulting luminosity-weighted average values and 1$\sigma$ confidence
intervals are $\mu_{\rm dust}=29.4 \pm 5.9$, $\mu_{\rm gas}=24.1\pm 4.8$ 
and $\mu_{\rm stars}=34.1\pm 6.8$, for the dust, gas and stellar component,
respectively. The variations are caused by small morphological 
differences between the couterparts and small positional offsets
between the dust and gas peaks of a few tenths of an arcsecond, less 
than the beam size. This shows that the impact of differential lensing 
between the gas and the dust is not larger than other systematic effects 
when deriving spatially-integrated results for the Emerald. The difference
in magnification between the dust and stellar components are also minor 
compared to the multiwavelength configurations of other high-redshift 
SMGs strongly lensed by galaxy clusters 
\citep[e.g.,][]{mackenzie14,timmons16}. Magnification factors change by
up to 30\% in our alternative models and are consistent with those
measured from light-traces-mass models in \citet{frye18}, suggesting 
that the remaining model degeneracies are not a major concern for 
the analysis of the Emerald.

We find $\mu = 3.8 \pm 0.5$ and $\mu = 6.1 \pm 0.9$ for the dust
continuum emission in the more compact submm sources south (Co-S) 
and north (Co-N) from the arc, respectively. These values become 
about 20\% lower when considering instead their gas emission.

\subsection{Source-plane reconstruction}
\label{ssec:sourceplanereconstruction}

In Figure~\ref{fig:sourceplane} we show the reconstructed source 
plane morphologies of the gas and dust in the Emerald. We used the
best-fitting lens model of the foreground cluster to infer the
intrinsic size and position of the source seen as an extended
arc. The source falls very near the critical line, so that the 
magnification varies by a factor of at least ten between the center 
and the two extreme ends (see inset in Fig.~\ref{fig:lensmodel}). 
For such a configuration, reconstructing correctly the source plane 
morphology of the dust and the molecular gas requires a specific
procedure, such as those described in more detail in \citet{johnson17} 
and \citet{fu12}. We followed a similar approach and described the 
source-plane profile of the gas and dust components with simple 
2D elliptical Gaussian models, computed the associated brightness 
intensity in the image plane, convolved with the beam, and compared 
with the PdBI and SMA images.

The CO(4--3) emission is modeled with six parameters in the source
plane, its position, semi-major and semi-minor axes, orientation 
and total flux density. We drew these parameters from Gaussian
distributions centered on the values derived from the MCMC 
calculations, and used {\sc Lenstool} to ray-trace the associated source
profile to the image plane through the best-fitting potential derived
in Sect.~4.2. We then convolved with the PdBI beam and computed the 
residual between the modeled and observed images. The $\chi^2$ was 
derived from the residual map, by considering all pixels that fall 
within the $3\sigma$ contours of the spectrally-integrated CO line 
emission. We explored the parameter space by iterating 1000 times, and 
adopted the source profile associated with the lowest $\chi^2$ in the 
image plane as the best fit. We performed a similar reconstruction of 
the {\it HST}/F110W emission. For the dust continuum, we assumed
a circular Gaussian profile given that the arc is not resolved in the 
tangential direction by the SMA beam. The resulting maps and source 
plane models are shown in Fig.~\ref{fig:sourceplane}.

The best-fit position of the stellar continuum emission is consistent
with that of the gas and dust centroids within the uncertainties. All 
components also have comparable projected spatial extents, resulting in 
a size of 2.7~kpc$~\times$~1.7~kpc in the source plane, elongated along 
position angle PA $=-73^\circ$ (east from north). Systematic uncertainties 
on these size estimates are up to about 25\%, according to our alternative 
lens models. The stellar continuum is spatially resolved along both the 
major and minor axes, so that these sizes are both intrinsic; however, 
the major axis of the gas and dust measurement are dominated by the beam 
size. It is nonetheless encouraging to find similar sizes, and we will 
in the following discussion assume that the dust, gas, and stellar 
components are distributed over similar regions in the source plane.

Since the \textit{HST} imaging resolves the rest-frame optical continuum 
of the Emerald, we can reconstruct the stellar morphology in the 
source plane by directly ray-tracing the observed WFC3/F110W
image pixel by pixel through the lensing mass distribution with the 
{\tt cleanlens} algorithm \citep{sharon12}; {\tt cleanlens} models 
the intrinsic morphology by allowing the source-plane pixels to have 
arbitrary distortions and sizes to match those of the unlensed source. 
The delensed stellar continuum in the Emerald has a very elongated 
shape, as expected given the reconstructed profile of the point-spread 
function of the \textit{HST}, which has an axis ratio of about 0.15. 

Finally, the source-plane reconstruction allows us to infer the relative
positions of the arc and the two additional galaxies Co-N and Co-S. The 
reconstruction shows that the intrinsic positions of the arc and Co-N are 
consistent within 1$\sigma$, for the best-fit and alternative models. This
suggests that Co-N is another image of the same galaxy as the one forming 
the arc (i.e. the Emerald). For our best-fit model, Co-S falls at a 
projected distance of about 18~kpc from the Emerald in the source plane 
and is likely another, interacting or merging galaxy. Alternative models 
suggest an offset of up to about 30~kpc. Although these models give worse 
fits to the lensing configuration, their results do provide a rough (and 
perhaps overly pessimistic) constraint on the systematic uncertainties 
that we might expect.

\section{Stellar population, molecular gas, dust, and star formation in the Emerald}
\label{sec:lemeraude}

\subsection{Intrinsic integrated properties}
\label{ssec:intrinsicproperties}

After characterizing the foreground environment along the line of sight 
toward PLCK\_G165.7+49.0 and the lensing configuration, we now discuss 
the properties of the Emerald, the main submm arc. The integrated 
properties are derived in \citetalias{canameras15}, and 
\citet{harrington18} also present an integrated CO(1--0) spectrum of this
source obtained with the Green Bank Telescope. However, neither analysis 
included a detailed lens model, so that the intrinsic properties of the 
Emerald could not be given. The interpretation of these results is 
also complicated by the presence of the two sources Co-S and Co-N within 
the 20\arcsec--30\arcsec\ beams of \textit{Herschel}/SPIRE and typical 
single-dish telescopes in the submillimeter and millimeter.

\begin{figure*}
\centering
\includegraphics[width=0.45\textwidth]{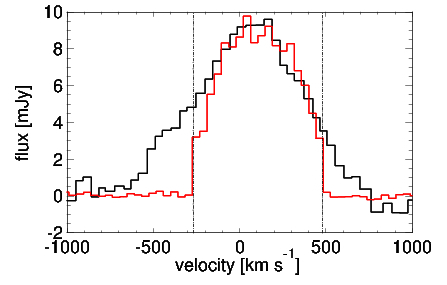}
\includegraphics[width=0.45\textwidth]{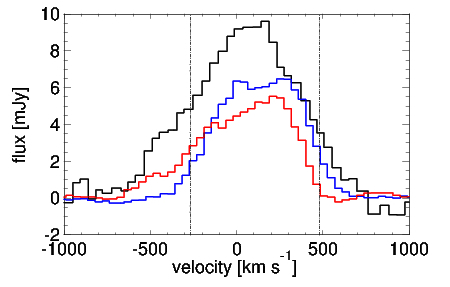}
\caption{PdBI CO(4--3) line profiles of the Emerald (the main arc of PLCK\_G165.7+49.0, 
red line in the {\it left} panel) and the two nearby sources, Co-N and Co-S 
(red and blue lines in the {\it right} panel), compared to the GBT CO(1--0) 
line profile of \citet[][black line in both panels]{harrington18}. The wings 
in the spatially-integrated CO(1--0) line profile are not detected 
toward the arc, but their velocity range is well-matched with the range 
covered by the CO(4--3) line emission from sources Co-N and Co-S, which are 
blended with the arc in the GBT beam.}
\label{fig:co10profiles}
\end{figure*}

\begin{table}
\centering
\begin{tabular}{lccc}
\hline
\hline
Quantity & Value & Unit & $\mu$ \\
\hline
$L_{\rm FIR,gb}$ & $(1.4 \pm 0.3) \times 10^{12}$ & L$_{\odot}$ & 29.4 $\pm$ 5.9 \\
$L_{\rm FIR,tpl}$ & $(1.8 \pm 0.4) \times 10^{12}$ & L$_{\odot}$ & 29.4 $\pm$ 5.9 \\
SFR & $176 \pm 35$ & M$_{\odot}$ yr$^{-1}$ & 29.4 $\pm$ 5.9 \\  
$M_{\rm d}$ & $(7.4 \pm 1.5) \times 10^7$ & M$_{\odot}$ & 29.4 $\pm$ 5.9 \\
$I_{\rm CO(4-3)}$ & 0.49 $\pm$ 0.10 & Jy km s$^{-1}$ & 24.1 $\pm$ 4.8 \\
$L_{\rm CO(4-3)}$ & $(2.4 \pm 0.5) \times 10^7$ & L$_{\odot} $ & 24.1 $\pm$ 4.8 \\
$L^\prime_{\rm CO(4-3)}$ & $(7.6 \pm 1.5) \times 10^9$ & K km s$^{-1}$ pc$^2$ & 24.1 $\pm$ 4.8 \\
$M_{\rm mol}$ & $(1.1 \pm 0.2) \times 10^{10}$ & M$_{\odot}$ & 24.1 $\pm$ 4.8 \\
\hline
\end{tabular}
\caption{Intrinsic dust, gas, and star formation properties of the Emerald, 
spatially-integrated over the submm arc, as inferred from the single-dish 
observations and analysis presented in \citetalias{canameras15} and 
Ca\~nameras et al. (2018, A\&A accepted). Each quantity 
is delensed using the relevant gravitational magnification factor, $\mu$, 
obtained in Sect.~\ref{ssec:magnificationfactors}, corrected for the fraction
of single-dish fluxes emitted by sources Co-N and Co-S, and divided by a
factor of 2 to account for the two merging images in the arc. Errors include 
statistical uncertainties on $\mu$. We list the FIR luminosities integrated 
over the 8--1000~$\mu$m range, for a simple graybody function and for 
mid-infrared-to-millimeter templates \citepalias[see][]{canameras15}.}
\label{tab:intrinsicproperties}
\end{table}

We used the lens modeling from Sect.~\ref{ssec:magnificationfactors} 
to derive intrinsic source properties from the observed integrated results. 
The total flux of the main arc and sources Co-N and Co-S measured on the SMA 
EXT map is $(71.6 \pm 0.6)$~mJy, corresponding to 80\% of the flux measured 
with SCUBA-2. Assuming that all sources have similar dust temperatures, 
our 850~$\mu$m SMA map suggests that $(81 \pm 1)$\% of the total FIR 
luminosity detected with \textit{Herschel} and the single-dish radio 
telescopes are emitted by the main arc, and $(12 \pm 1)$\% and $(7 \pm 1)$\% 
by sources Co-N and Co-S, respectively. Accounting for the magnification 
factors derived in Sect.~\ref{ssec:magnificationfactors}, this results in 
intrinsic star-formation rates of SFR$_{\rm arc}=(176 \pm 35)$~M$_{\odot}$~yr$^{-1}$ 
for the main arc, and SFR$_{\rm Co-N}=(252 \pm 39)$~M$_{\odot}$~yr$^{-1}$ and 
SFR$_{\rm Co-S}=(236 \pm 28)$~M$_{\odot}$~yr$^{-1}$ for the compact sources.
We obtain a lower delensed star-formation rate for the Emerald compared to 
Co-N likely because, due to the parity inversion, none of the two merging 
images forming the submm arc are complete images of the intrinsic source. 
These star-formation rates are given for a \citet{chabrier03} stellar 
initial mass function (IMF), and are therefore a factor~1.8 lower than 
when applying the popular prescription by \citet{kennicutt89}. 
Table~\ref{tab:intrinsicproperties} gives a summary of the intrinsic 
spatially-integrated dust, gas, and star formation properties of the 
Emerald inferred from the observational results of \citetalias{canameras15} 
and Ca\~nameras et al. (2018, A\&A accepted), by correcting for the 
strong lensing magnification and by a factor of two to account for the 
two images forming the arc.

The Emerald is a ULIRG with an intrinsic far-infrared luminosity,
$L_{\rm FIR}= 1.8 \times 10^{12}$~L$_{\odot}$, and intrinsic dust and
molecular gas masses of $7.4 \times 10^7$~M$_{\odot}$ and $1.1 \times
10^{10}$~M$_{\odot}$, respectively. The latter value assumes a ``ULIRG''
conversion factor between CO luminosity and molecular gas mass of
$0.8~{\rm M}_{\odot} / ({\rm K\ km\ s}^{-1}\ {\rm pc}^2)$, and a line
luminosity ratio $R_{4,1}=L^\prime_{\rm CO(4-3)}/L^\prime_{\rm CO(1-0)}=0.55$, 
as found empirically by, for example, \citet{spilker14} and 
\citet{danielson11}. These masses are fully consistent with those 
measured for many submillimeter galaxies in the field 
\citep[e.g.,][]{tacconi08}, and the Emerald is thus a representative 
member of this class of galaxy.

A significantly lower ratio, $R_{4,1}=0.25 \pm 0.10$, is found if we use 
the integrated CO(4--3) line luminosity we measured with the IRAM 30-m 
telescope, $\mu L^\prime_{\rm CO(4-3)}=(46.0 \pm 3.1) \times 10^{10}$~K km 
s$^{-1}$ pc$^2$ (Ca\~nameras et al. 2018, A\&A accepted), and the CO(1--0) 
line luminosity from \citet{harrington18}, who recently detected this line 
with the Green Bank Telescope (GBT) for a few lens candidates drawn from 
the \textit{Planck} all-sky survey, including five GEMS. Their CO(1--0) 
flux estimates are overall unusually high compared to other high-redshift 
galaxies, and highlight that this also results in exceptionally high gas 
masses, high gas-to-dust ratios, and low gas excitations, akin to local 
ULIRGs rather than other high-redshift galaxies (in spite of their high 
star-formation intensities).

In Figure~\ref{fig:co10profiles} we compare the \citet{harrington18} CO(1--0) 
line profile with the PdBI CO(4--3) line profile integrated over the arc. We 
find significant differences, with about 30\% of the CO(1--0) flux emitted
at velocities outside the range covered by CO(4--3); this is not expected 
if both lines trace the same clouds. The velocity range in the wings is 
well-matched with the velocity range of the sources Co-N and Co-S, which 
also fall within the FWHM~$\simeq$~21\arcsec\ beam of the Green Bank 
Telescope at about 36~GHz, the redshifted frequency of CO(1--0) in the 
Emerald.

If we subtract the wings of the CO(1--0) line from the GBT spectrum, and 
correct for missing flux in our PdBI spectrum spatially-integrated over the 
arc, we find $R_{4,1} \simeq 0.45$, comparable to the typical values found 
by \citet{spilker14} and \citet{danielson11}. However, this higher value 
is also a lower limit to the intrinsic ratio in the Emerald, because of 
contamination with Co-N and Co-S. The source confusion within the beam 
of the GBT therefore prevents an in-detail comparison between the 
$J=1$--0 and $J=4$--3 CO line emissions and motivates our choice of 
adopting a fiducial value of $R_{4,1}=0.55$.

\subsection{Dust and gas morphology and gas kinematics}
\label{ssec:starformation}

\begin{figure*}
\includegraphics[width=0.95\textwidth]{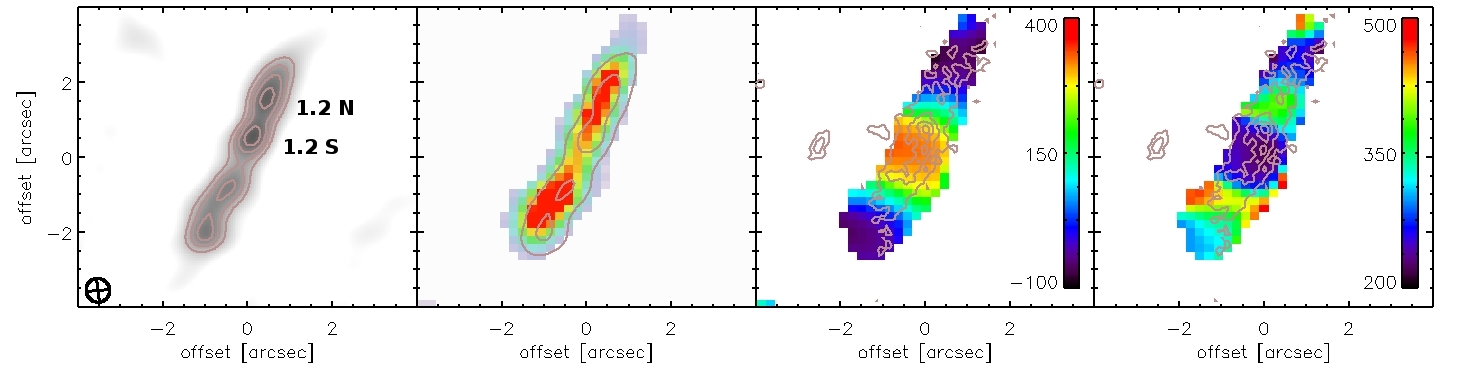}
\includegraphics[width=0.95\textwidth]{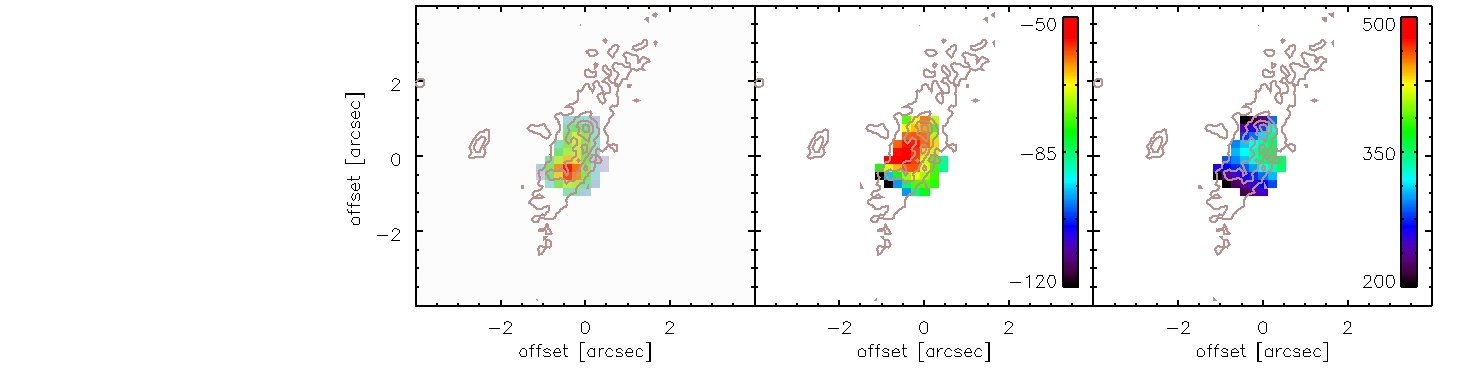}
\caption{{\it Top:} Dust and gas morphology and gas kinematics of the Emerald, 
the main submm arc of PLCK\_G165.7+49.0, from single-component Gaussian 
line fits. {\it Left to right:} Maps of the SMA 850~$\mu$m dust and PdBI 
CO(4--3) morphologies, relative velocity, and Gaussian line FWHM. 
The arc is composed of two multiple images of the same source, \#1.1 on the 
southern portion and \#1.2 on the northern one, respectively (see also 
Fig.~\ref{fig:lensmodel}). We identify two star-forming clumps per image
and focus most of the analysis on the two northern ones, called \#1.2~S and 
\#1.2~N. The SMA and PdBI beam FWHMs are 0.90\arcsec~$\times$~0.75\arcsec 
and 0.76\arcsec~$\times$~0.75\arcsec, respectively. Velocity offsets and 
line widths are given in km~s$^{-1}$. Contours on the two leftmost panels
show the SMA 850~$\mu$m dust continuum, starting at +6$\sigma$ and increasing
in steps of +2$\sigma$, and those on other panels show the stellar continuum 
in the F110W band from \textit{HST}/WFC3. {\it Bottom, left to right:} 
Same maps for the secondary Gaussian component detected at $\geq 3 \sigma$ 
over a region of about 2.5\arcsec~$\times$~1.5\arcsec\ near the center of 
the arc.}
\label{fig:comaps}
\end{figure*}

We show the SMA 850~$\mu$m morphology and the CO morphology and kinematics 
of the Emerald in Fig.~\ref{fig:comaps}. The long-wavelength emission is 
associated with the central regions of the western arc of 28.5\arcsec\ shown 
in Fig.~\ref{fig:panorama}. The gas and dust emissions are extended along
an arc of about 5.3\arcsec\ length, which is not spatially resolved along 
the direction parallel to the critical line (and perpendicular to the 
magnification axis). The FWHM size along the minor axis is 0.84\arcsec, 
compared to a beam size of 0.76\arcsec~$\times$~0.75\arcsec. The Emerald 
is composed of four clumps in total, with the two southernmost clumps being 
strongly blended and difficult to separate. This is better seen at the 
somewhat higher resolution of the CO than the dust image. Within the beam 
size, the gas and dust morphologies are broadly consistent with each other,
except for a small, roughly 0.1\arcsec\ positional offset for some clumps, 
which is much less than the beam size in either image.

About 60\% of the dust emission originates from the clumps, the rest from 
a fainter component between. It is difficult to constrain whether this 
component is also clumpy or more diffuse because the emission is faint and 
blended with the clumps, and the intraclump regions are spatially not well 
resolved.

The velocity structure of the arc is seen in our PdBI CO(4--3) maps
(Fig.~\ref{fig:comaps}), which we constructed in the way described in 
Sect.~\ref{ssec:obs.iram}. Relative velocities increase from southeast 
to northwest from $-63 \pm 15$~km~s$^{-1}$ to $+336 \pm 15$~km~s$^{-1}$, 
and then decrease again to $-70 \pm 15$~km~s$^{-1}$. Gaussian line widths, 
$\sigma$, are rather moderate and range from $93 \pm 25$~km~s$^{-1}$ 
to $190 \pm 25$~km~s$^{-1}$; they are shown in the right panel of
Fig.~\ref{fig:comaps}. It is interesting that the line widths in the 
center of the arc, near the critical line of the best-fitting lensing 
model, are more narrow than further out. If several, partially 
overlapping images were present we would expect the line widths to 
increase because of blending, and if the velocity offsets across the 
major axis were indicating a merger of two rotating gaseous disks, we 
would also expect a higher turbulence and broader line profiles at 
this position. We therefore consider the narrow lines as additional, 
supporting evidence that the arc comprises two lensed images of the 
same region, with a parity inversion on both sides of the critical 
line.

We can estimate a dynamical mass from the velocity gradient, $v$, of
380~km~s$^{-1}$ and the intrinsic FWHM size of the Emerald of 2.7~kpc  
(Sect.~\ref{ssec:sourceplanereconstruction}). If we assume, as is
usually done, that the gradient encompasses both sides of a disk, then
we find a dynamical mass of $M_{\rm dyn} = (v/\sin{i})^2 R / G = 1.1 \times
10^{10}$~M$_{\odot}$, where $R$ is the disk radius, and $G$ the
gravitational constant. This estimate ignores possible inclinations, 
$i$, or a mismatch between lensing and the rotational major axis, which 
may lead to incomplete sampling of the rotation curve. The resulting 
mass is much lower than the sum of the gas and stellar masses of 
$5.1 \times 10^{10}$~M$_{\odot}$ (Sections~\ref{ssec:intrinsicproperties} 
and \ref{ssec:stellarpop}), implying that we are either observing a disk 
nearly seen face-on at $\la 20^\circ$ inclination angle, a disk magnified 
approximately along the kinematic minor axis, or a disk where only 
one side of the rotation curve is being magnified. In the latter case, 
we obtain $M_{\rm dyn} = 9.1 \times 10^{10}$~M$_{\odot}$, by setting velocity 
$v=380$~km~s$^{-1}$ and $R=2.7$~kpc, about 40\% larger than the baryonic 
mass. The discrepancy would be alleviated if we had used a \citet{salpeter55} 
initial mass function instead of the typically adopted \citet{chabrier03} 
IMF, and as favored by the lens of another GEMS, the Ruby 
\citep[][\citetalias{canameras17b} hereafter]{canameras17b}. Neither 
mass estimate, however, places the Emerald outside the typical mass 
range of massive, dusty, intensely star-forming galaxies at $z\sim2$
\citep[e.g.,][]{casey14}.

\subsubsection{Systematic effects of lensing on the kinematic measurements}

Strong gravitational lensing near the critical line affects the 
brightness distribution of the gas and dust emission, and
obviously plays a large role in deriving integrated continuum and
line fluxes and related quantities. Integrated line profiles and
velocity offsets, however, are a convolution of the intrinsic
kinematics and emission-line surface-brightness distribution, and
might therefore also be affected by the details of the lens
reconstrution, and could potentially add important systematic
uncertainties to estimates of intrinsic source properties. To
investigate the impact of gravitational lensing on the emission-line
parameters, we have extracted the spectra within three apertures:
including the region showing a blue emission-line component in 
Fig.~\ref{fig:comaps}; a 0.6\arcsec\ wide annulus around that
region; and the remainder of the Emerald, before and after our 
pixel-by-pixel correction for the gravitational magnification. We 
then compared the results of our Gaussian line fitting in each case.

The difference of the kinematic properties of each line component in
the image and source plane are in fact very small. When extracting the
spectra without the lensing correction, we find relative velocities of
318~km~s$^{-1}$, 84~km~s$^{-1}$, 161~km~s$^{-1}$, and --7~km~s$^{-1}$
for the narrow and broad components associated with the clump, the 
spectra from the annulus and the remainder of the Emerald, respectively. 
With the exception of the annulus, these are within 10~km~s$^{-1}$ of 
the values measured from the spectra after correcting for the local 
magnification factors. The FWHM line widths are 218~km~s$^{-1}$, 
418~km~s$^{-1}$, 443~km~s$^{-1}$, and 360~km~s$^{-1}$, even closer to 
those measured after the correction. This suggests that the impact of 
gravitational lensing on the derived emission-line kinematics is 
negligible, at least for our analysis of the Emerald.

\subsubsection{Systematics on the gas kinematics from clump identification}

As a final test of our analysis of the resolved dust and CO line
emission in the Emerald, we also used {\sc Clumpfind} 
\citep[][]{williams94} to quantify the surface-brightness distribution
and kinematic substructure in the arc of the Emerald in a more
reproducible way than the one used when identifying these structures by
eye in a rather heuristic way. {\sc Clumpfind} is a publicly available
IDL-routine that identifies contiguous structures within an imaging
spectroscopy data cube, starting with the brightest peak in the cube,
and then lowering the flux threshold with a step size that can be
selected by the user. We started at 11.8~mJy beam$^{-1}$ and decreased
the flux in steps of 1.8~mJy beam$^{-1}$ (3~$\times$ the rms of our
data cube), until we reached a threshold of 3~mJy beam$^{-1}$
(corresponding to 5$\sigma$). 

{\sc Clumpfind} finds the same two extended components, which we also 
identified by eye, and in addition, the secondary, blueshifted component 
in the center of the source, which it identifies as as single structure. 
In addition to the structure within the arc, {\sc Clumpfind} also 
identifies the two separate sources, Co-N and Co-S, each associated with 
a single clump. This confirms our by-eye analysis. All individual 
clumps are detected at ${\rm SNR=12}$ and 20 per beam in the CO line 
emission and dust continuum, respectively, well above the signal-to-noise 
ratios at which \citet{hodge16} found spurious clumpiness in high-resolution 
dust imaging of high-redshift galaxies with ALMA.

\subsection{Resolved stellar component}
\label{ssec:stellarpop}

\begin{figure}
\centering
\includegraphics[width=0.4\textwidth]{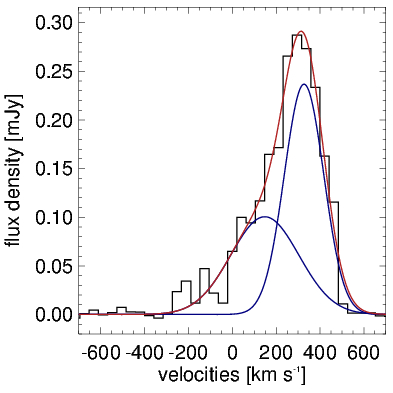}
\caption{Spatially-integrated CO(4--3) spectrum of the region shown 
in Fig.~\ref{fig:comaps} where we detect a secondary emission-line 
component. The red line shows the double-Gaussian fit to the overall 
spectrum and blue lines show the individual blueshifted and systemic 
components.}
\label{fig:windspec}
\end{figure}

\begin{table}
\centering
\begin{tabular}{lcc}
\hline
\hline
Band & Flux density & Unit \\
\hline
CFHT/MEGACAM $r$ & $<$8.5 & nJy \\
CFHT/MEGACAM $z$ & $<$29.3 & nJy \\
\textit{HST}/WFC3 F110W & 47.8 $\pm$ 3.8 & nJy \\
CFHT/WIRCAM $J$ & 64.5 $\pm$ 5.3 & nJy \\
\textit{HST}/WFC3 F160W & 110 $\pm$ 9 & nJy \\
CFHT/WIRCAM $K_{\rm s}$ & 345 $\pm$ 13 & nJy \\
\textit{Spitzer}/IRAC 3.6 $\mu$m & 1.19 $\pm$ 0.01 & $\mu$Jy \\
\textit{Spitzer}/IRAC 4.5 $\mu$m & 1.86 $\pm$ 0.02 & $\mu$Jy \\
WISE W3 & 9.5 $\pm$ 2.7 & $\mu$Jy \\
WISE W4	& 204 $\pm$ 19 & $\mu$Jy \\
\textit{Herschel}/SPIRE 250 $\mu$m & 13.4 $\pm$ 0.3 & mJy \\
\textit{Herschel}/SPIRE 350 $\mu$m & 11.6 $\pm$ 0.2 & mJy \\
\textit{Herschel}/SPIRE 500 $\mu$m & 7.3 $\pm$ 0.2 & mJy \\
JCMT/SCUBA-2 850 $\mu$m & 1.4 $\pm$ 0.4 & mJy \\
IRAM/GISMO 2~mm & 123 $\pm$ 39 & $\mu$Jy \\
\hline
\end{tabular}
\caption{Intrinsic photometry of the Emerald obtained from our new 
optical/near-infrared imaging with the CFHT, {\it HST} and {\it Spitzer}, 
and by correcting the source-integrated (sub-)millimeter flux densities 
presented in \citetalias{canameras15} for the contribution of Co-N and 
Co-S. We demagnified the optical and near-infrared stellar continuum fluxes 
by the gravitational magnification factor $\mu_{\rm stars}=34.1 \pm 6.8$ and 
the dust continuum fluxes by $\mu_{\rm dust}=29.4 \pm 5.9$, both derived 
from our best-fitting lensing model. Since the stellar and dust components 
likely have similar contibutions to the observed WISE emission, we used
the average value of $\mu$ to correct the fluxes in the W3 and W4 bands
\citep[as done in][]{timmons16}. We also give the 3$\sigma$ upper limits 
of non-detection in MEGACAM $r$- and $z$-bands.}
\label{tab:photo}
\end{table}

The near-infrared \textit{HST}/WFC3 emission from the western arc
extends over a total length of 28.5\arcsec, and consists of at least 
12 multiple images. We focus again on the Emerald, the central 
region of the arc, also detected in CO and 850~$\mu$m dust emission. 
Its stellar continuum morphology is best seen in the bottom panel of 
Fig.~\ref{fig:sourceplane}. In the rest-frame optical this component
consists mainly of diffuse emission, with a bright clump near the 
center of the arc, where the gravitational magnification is greatest. 
The intrinsic FWHM size of this region is 2.7~kpc~$\times$~1.7~kpc 
along the major and minor axis, respectively, using our lensing model 
presented in Sect.~\ref{ssec:magnificationfactors}. The size of the 
arc in the WFC3 images is comparable to that seen in the dust and gas
(Sect.~\ref{ssec:starformation}), but the morphologies are not
strictly the same, since the stellar continuum reaches the highest
surface-brightnesses where the gas and dust emissions are faintest. 
In turn, the brightest clumps seen at long wavelengths correspond to
rather faint regions in the stellar continuum. It is entirely possible
that this is mainly a sign of variations in dust and cloud cover.

We probed the intrinsic stellar continuum properties of the Emerald
using the counterparts of the submm arc in our CFHT, WFC3 and IRAC
imaging. To correct for some faint underlying continuum emission from
member galaxies of the foreground cluster, we modeled the three lens 
galaxies in the western group with S\'ersic profiles using {\sc Galfit}
\citep{peng10}. The deblended fluxes of the high-redshift 
source are measured in the lens-subtracted residual images, with a 
prior in the $K_{\rm s}$-band. These fluxes are then divided by a factor 
of two, to account for the two merging images producing the arc, and 
corrected for the magnification factor $\mu_{\rm stars}$ reported above. 
We also corrected the SPIRE, SCUBA-2 and IRAM 30-m/GISMO single-dish flux 
densities for gravitational lensing using $\mu_{\rm dust}$, divided by a 
factor of two to account for the two blended counter images, and corrected
for the roughly 20\% of the total flux from sources Co-N and Co-S. 
The resulting optical to submillimeter photometry is presented in 
Table~\ref{tab:photo}.

Using the simple stellar population models (SSPs) from \citet{bruzual03}, 
with solar metallicity, a \citet{chabrier03} stellar IMF, and the 
\citet{calzetti94} extinction law, we obtain a best-fitting model for 
very young stellar populations of about 25~Myr and $A_{\rm V} \simeq 2.7$~mag, 
with a goodness-of-fit of $\chi^2=2.7$. We also obtain good fitting 
results when using exponentially declining star-formation histories 
instead, with ages below 50~Myr.

We then used {\sc Magphys} \citep{dacunha08,dacunha15} to better
constrain the amount of energy reprocessed by dust through our
long-wavelength photometry, and to derive a robust stellar mass
estimate for the Emerald. {\sc Magphys} has proven to give consistent 
results with other codes using different template libraries
\citep[e.g.,][]{nayyeri17}. A possible AGN contribution to the dust
heating is neglected, but this is not a concern here, since we have
already shown that AGN heating does not dominate the far-infrared
spectral energy distribution \citepalias{canameras15}.

\begin{figure}
\centering
\includegraphics[width=0.5\textwidth]{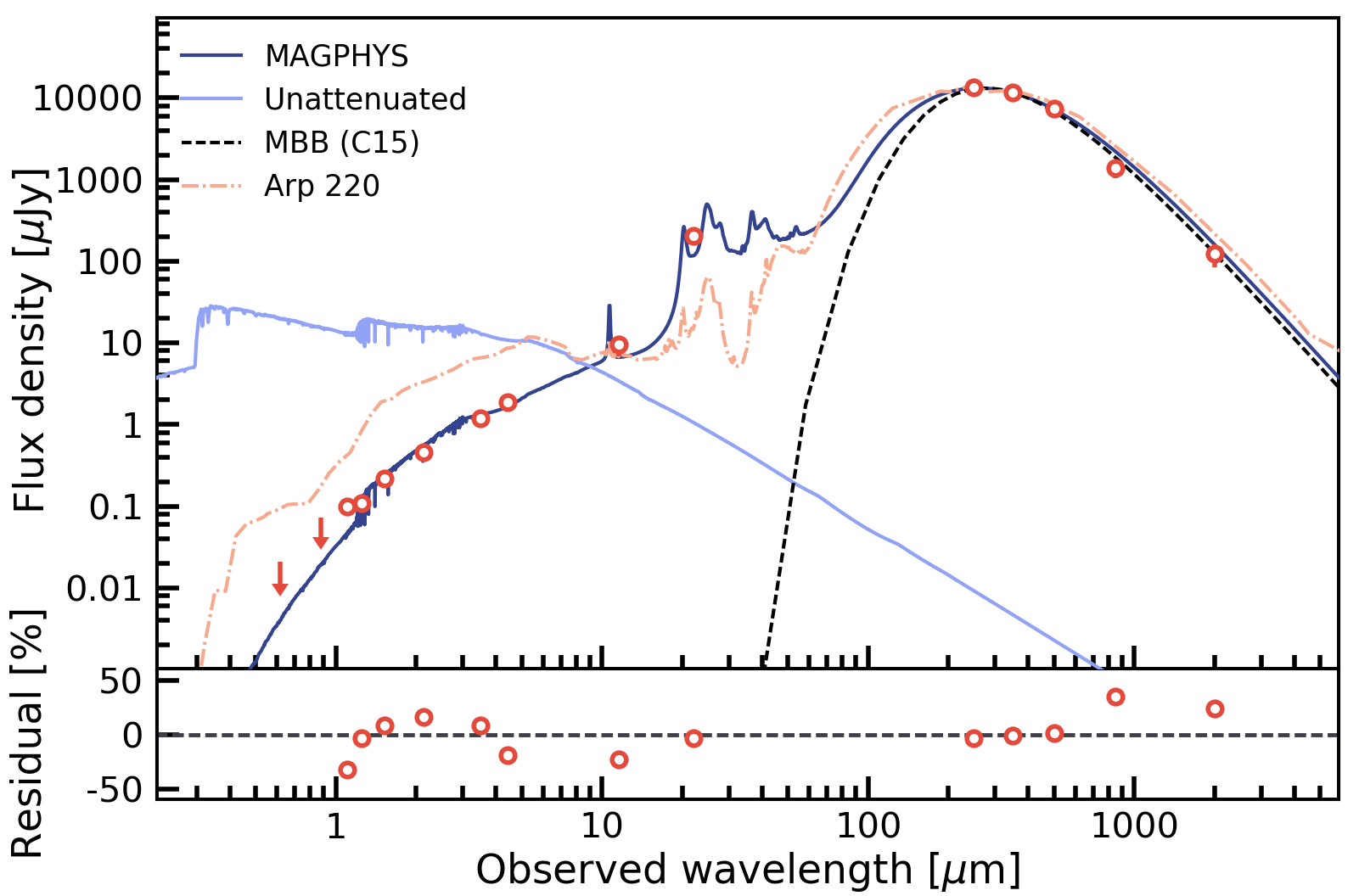}
\caption{{\it Top:} Intrinsic spectral energy distribution of the Emerald
from the optical to the millimeter. Red circles show the photometry 
of the arc presented in Table~\ref{tab:photo}, corrected for the 
gravitational magnification factors of the stellar and dust components,
as described in the text. Flux uncertainties are smaller than the 
symbols and downward arrows show the $3\sigma$ upper limits on the 
MEGACAM fluxes. The best-fit SED to the full wavelength range obtained 
with {\sc Magphys} is plotted as a solid blue curve, and the best-fit 
stellar continuum without dust attenuation is shown in light blue. 
The dash-dotted orange line indicates the best-fit template of the 
local starburst galaxy Arp~220, shifted to $z=2.236$ and normalized 
to the flux density of the Emerald in the 350~$\mu$m band of SPIRE. 
{\it Bottom:} Residuals of the best-fitting {\sc Magphys} model.}
\label{fig:magphys}
\end{figure}

We fit the delensed SED of the arc with the values listed in
Table~\ref{tab:photo}, assuming a redshift $z=2.236$, and obtain the
best-fit SED shown in Fig.~\ref{fig:magphys}, with a goodness-of-fit
of $\chi^2=3$. The resulting instrinsic stellar mass of the source 
that produces the submm arc is $M_* = (4.1 \pm 0.4) \times 10^{10}$~M$_{\odot}$, 
for a Chabrier IMF, with $A_{\rm V}=3.9 \pm 0.3$~mag, and an intrinsic 
star-formation rate (SFR) of $(74 \pm 3)$~M$_{\odot}$~yr$^{-1}$
(corrected for a gravitational magnification factor of 
$\mu=34.1$). The corresponding stellar mass surface density is 
$(11 \pm 5) \times 10^9$~M$_{\odot}$ kpc$^{-2}$ for an intrinsic size of 
2.7~kpc~$\times$~1.7~kpc along the (delensed) major and minor axis,
respectively. The dust properties inferred with {\sc Magphys} are
consistent with those listed in Table~2 of \citetalias{canameras15},
after correcting for the gravitational magnification where necessary.

The blue clump very near to the critical line, which may either be 
part of the Emerald or be an intervening source (see
Sect.~\ref{ssec:foregroundmassdistribution} and Fig.~\ref{fig:panorama}), 
contributes only marginally to the integrated flux of the Emerald. 
We do not include it in the SED fitting presented here, but we did run 
an alternative model which includes the clump, and found that the fit 
results remain within the 1$\sigma$ uncertainties. We also checked 
that the small, spatial offset between the rest-frame optical and 
submillimeter and millimeter position of the Emerald 
(Sect.~\ref{ssec:sourceplanereconstruction}) have no significant 
impact on the SED modeling with {\sc Magphys}, by reproducing the fit 
without the WFC3 and CFHT fluxes. The results are consistent with those 
of the best-fit model within the 1$\sigma$ uncertainties, likely 
because the stellar mass is mainly constrained by IRAC photometry
in the rest-frame NIR.

The total star-formation rate found from the stellar continuum of
SFR$_{\rm opt}=74$~M$_{\odot}$~yr$^{-1}$ is much lower than that obtained
from the far-infrared spectral energy distribution, SFR$_{\rm FIR}=176$
M$_{\odot}$ yr$^{-1}$, after correcting for a magnification factor of
$\mu = 29.4$ (see Sect.~\ref{ssec:magnificationfactors}). This
may indicate that most of the star formation in the Emerald is
hidden behind high dust and gas column densities. This is also
consistent with the morphologies of the dust, gas, and stellar
component. High-surface brightness dust emission extends over much
larger radii than the bright stellar continuum (although faint
continuum emission is also probed over larger scales). Given that the
stellar component falls very near the caustic line, and that the
overall distribution of gas and dust in intense star-forming regions
is likely very clumpy \citep[e.g.,][]{genzel12,swinbank15,iono16}, it 
seems plausible that we are fortuitously seeing along a relatively 
low-extinction sight-line into the starburst.

\subsection{Additional sources}
\label{sec:othergalaxies}

\begin{table*}
\begin{tiny}
\begin{tabular}{lcccccccccc}
\hline
\hline
Source & RA & Dec & $D_{\rm maj} \times D_{\rm min}$ & PA & $S_{\rm 850}$ & $L_{\rm FIR}$ & SFR & $I_{\rm CO(4-3)}$ & $L^\prime_{\rm CO(4-3)}$ & $M_{\rm mol}$ \\
 & (J2000) & (J2000) & [arcsec$^2$] & [deg] & [mJy] & [$10^{12}$ L$_{\odot}$] & [M$_{\odot}$ yr$^{-1}$] & [Jy km s$^{-1}$]& [$10^{10}$ K km s$^{-1}$ pc$^2$] & [$10^{10}$ M$_{\odot}$] \\
\hline
Co-N & 11:27:13.85 & $+$42:28:16.56 & 1.1 $\times$ 0.9 & 60 $\pm$ 2 & 1.8 $\pm$ 0.3 & 2.6 $\pm$ 0.4 & 252 $\pm$ 39 & 0.72 $\pm$ 0.16 & 1.1 $\pm$ 0.2 & 1.6 $\pm$ 0.3 \\ 
Co-S & 11:27:13.89 & $+$42:28:35.30 & 0.9 $\times$ 0.9 & 44 $\pm$ 2 & 1.7 $\pm$ 0.2 & 2.5 $\pm$ 0.3 & 236 $\pm$ 28 & 0.67 $\pm$ 0.13 & 1.1 $\pm$ 0.2 & 1.5 $\pm$ 0.3 \\
\hline
\end{tabular} 
\end{tiny}
\caption{Properties of compact components Co-N and Co-S. Columns are: source name; right 
ascension and declination; apparent major axis and minor axis FWHM, and position 
angle measured on the PdBI CO(4--3) map; intrinsic flux density at 850~$\mu$m; 
delensed FIR luminosity integrated in the range 8--1000~$\mu$m and star-formation 
rate given for a Chabrier IMF; intrinsic velocity-integrated flux and luminosity 
of CO(4--3) line; molecular gas mass assuming $R_{4,1}=0.55$ and 
$\alpha_{\rm CO}=0.8$~M$_{\odot}$/(K~km s$^{-1}$ pc$^2$). Intrinsic quantities are 
corrected for the local magnification factors, $\mu_{\rm Co-N}=6.1 \pm 0.9$ and 
$\mu_{\rm Co-S}=3.8 \pm 0.5$, inferred in Sect.~\ref{ssec:magnificationfactors}, 
and errors include statistical uncertainties on $\mu$.}
\label{tab:othergalaxies}
\end{table*}

\begin{figure}
\centering
\includegraphics[width=0.4\textwidth]{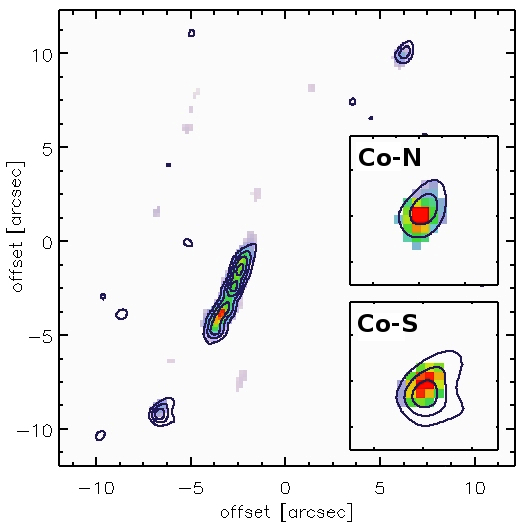}
\caption{Dust and gas morphology of the Emerald and the two neighboring 
compact sources Co-S and Co-N toward southeast and northwest, respectively.
The two insets have sizes of 3\arcsec~$\times$~3\arcsec. The color scale 
displays the PdBI CO(4--3) emission and the contours show the SMA 850~$\mu$m 
dust continuum.}
\label{fig:compact}
\end{figure}

As mentioned previously, the Emerald is surrounded by the two neighboring 
sources Co-S and Co-N at projected distances of 5\arcsec\ and 13\arcsec\ 
toward southeast and northwest, respectively (see Fig.~\ref{fig:compact}). 
These components were first introduced in Sect.~\ref{ssec:sourceplanereconstruction}, 
where we also listed their luminosity-weighted gravitational magnification 
factors, $\mu_{\rm Co-N} = 6.1 \pm 0.9$ and $\mu_{\rm Co-S} = 3.8 \pm 0.5$. The 
two sources are almost compact, with projected major axis length of about 
1.1\arcsec\ and 0.9\arcsec, respectively, in the CO line image obtained with 
the IRAM interferometer, for a beam size of 0.76\arcsec~$\times$~0.75\arcsec. 
Our lensing model suggests that source Co-N is an additional counter-image of 
the same galaxy that also gives rise to the Emerald (seen with a much lower 
magnification factor), whereas Co-S is a distinct component in the source 
plane. Our best-fitting lensing model favors a scenario where Co-S is a 
companion galaxy about~18~kpc away from the Emerald, although the relative 
projected distance is uncertain given the remaining degeneracies in the model. 
Major and minor axis sizes in the source plane are 5.9~kpc~$\times$~3.1~kpc 
for Co-S and 6.0~kpc~$\times$~2.7~kpc for Co-N.

We show the integrated spectra of both galaxies in Fig.~\ref{fig:co10profiles}. 
Their intrinsic properties are listed in Table~\ref{tab:othergalaxies}, and 
were derived with the same methods and assumptions as previously used for the 
Emerald. Values found for Co-N are consistent with those for the Emerald itself, 
further suggesting that we may be seeing another image of the same galaxy with 
a smaller magnification factor.

\subsection{Star-formation law in the Emerald}
\label{ssec:schmidtkennicutt}

Star-formation surface densities (star-formation intensities) and gas-mass 
surface densities are closely related through the Schmidt-Kennicutt 
relationship \citep[][]{schmidt59,kennicutt89}, and allow 
us to characterize the star-formation processes in galaxies over more 
than six orders of magnitude in star-formation rate and gas density. 
The position of a galaxy within this diagram is an indicator of the
efficiency with which gas is being turned into stars. While the main
physical driver (or drivers) of this relationship are still a matter
of active debate, different scenarios of how star formation is being
regulated in these galaxies make different predictions for where a
galaxy should fall within this model (which might not, however,
necessarily be unique). 

\subsubsection{Star-formation intensities}

In order to investigate the star formation properties of the Emerald with
the aid of the Schmidt-Kennicutt diagram, we need to translate our
spatially resolved dust and CO surface brightnesses into star-formation 
intensities and gas-mass surface densities. This requires several 
assumptions. First, we obtain star-formation intensities by assuming 
that all the dust within the Emerald has a single temperature, which 
corresponds to the value $T_{\rm d}=42.5$~K measured in 
\citetalias{canameras15} from the spatially-integrated FIR-to-millimeter
spectral energy distribution. This allows us to convert the local surface
brightnesses of the 850~$\mu$m continuum into local star-formation
intensities, by setting SFR $=2.5 \times 10^{-44} L_{\rm FIR}$, where SFR 
is given in M$_{\odot}$~yr$^{-1}$. The far-infrared luminosity, $L_{\rm FIR}$, 
is in erg~s$^{-1}$ and $L_{\rm FIR}$ is integrated over 8-1000~$\mu$m
\citep[][]{kennicutt89}. We remind the reader that this estimate is 
appropriate for a Chabrier stellar IMF and is a factor~1.8 (0.26~dex) 
lower than that originally adopted by \citet{kennicutt89}. The 
\citet{chabrier03} parametrization is currently the most commonly adopted 
in high-redshift studies, and we adopted this IMF to be consistent with 
these studies. We thus correct all values in this section and those 
adopted from the literature to take this into account, while understanding
that this might introduce a bias, since massive galaxies may be better 
characterized by a \citet[][]{salpeter55} IMF \citep[e.g.,][]{canameras17a}.

This approach results in central star-formation rate surface densities
between 38 and 45~M$_{\odot}$~yr$^{-1}$~kpc$^{-2}$ in the four clumps seen 
in the dust continuum, in images \#1.1 and \#1.2 of the Emerald
(Table~\ref{tab:clumps}). Since lensing conserves surface brightness, 
we did not need to correct these quantities for the magnification factors.

\subsubsection{Gas-mass surface densities}

The second quantity relevant for our analysis is the molecular gas mass 
surface density. We used the measured CO(4--3) fluxes, and followed
\citet{solomon97} to convert them into mass surface densities of molecular
hydrogen. To do so, we also needed to adopt a CO-to-H$_2$ conversion factor, 
a quantity whose value is still controversial. We adopted the ``standard''
conversion factor appropriate for ULIRGs \citep[][]{downes98},
$\alpha_{\rm CO}=0.8$~M$_{\odot}$/(K km s$^{-1}$ pc$^2$), which is
commonly used for intense starburst galaxies at high redshift and also 
consistent with our analysis of the gas-to-dust ratios in the GEMS 
\citepalias[see][]{canameras15}.

\begin{figure*}
\centering
\includegraphics[width=0.65\textwidth]{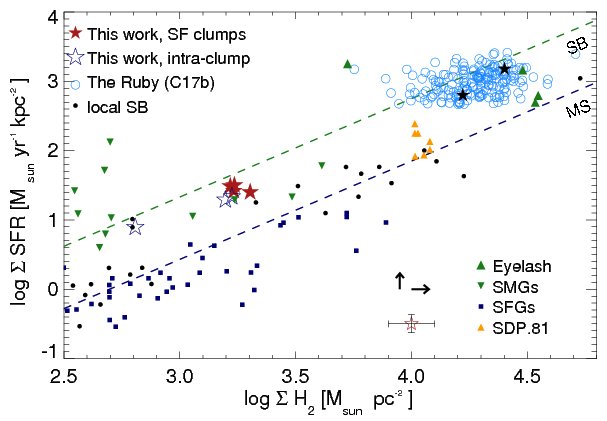}
\caption{Spatially-resolved Schmidt-Kennicutt diagram in the Emerald, for 
the four star-forming clumps identified with {\sc Clumpfind} (red stars) 
and the intraclump regions (open blue stars), the spatially resolved 
pixel-by-pixel analysis of the Ruby 
\citep[light blue circles,][]{canameras17b} and its luminosity-weighted
properties (black stars), and for other galaxy samples taken from the 
literature. These samples include spatially-resolved studies of the 
Eyelash \citep[green triangles,][]{swinbank11} and SDP.81 
\citep[yellow triangles,][]{hatsukade15}, submillimeter galaxies at
$z \sim 2$ \citep[green upside-down triangles,][]{bothwell10}, 
normal star-forming galaxies at $z = 1$--2.3 
\citep[blue squares,][]{tacconi10},
and local starbursts \citep[black circles,][]{kennicutt98}. The dashed 
lines labeled ``SB'' and ``MS'' show the ridge lines of ``starburst'' 
and ``main-sequence'' galaxies of \citet[][]{daddi10}, respectively 
\citep[see also][]{genzel10}. Typical error bars are shown in the 
lower-right corner. The small horizontal and vertical arrows indicate 
the expected offsets when using $R_{4,1}=0.45$ instead of 0.55, and the 
Salpeter stellar IMF instead of the Chabrier IMF, respectively (see 
text for details).}
\label{fig:sk}
\end{figure*}

Another complication is that we used a mid-$J$ transition of CO(4--3),
whereas the Schmidt-Kennicutt law was calibrated on CO(1--0). However,
emission from the ground rotational state of CO can be significantly 
contaminated by diffuse molecular gas outside of star-forming clouds 
\citep[][]{ivison11}. A possible caveat of directly using CO(1--0) 
measurements is therefore that the total line emission might overestimate 
the gas-mass surface densities within the star-forming regions themselves. 
As already discussed in Section~\ref{ssec:intrinsicproperties}, we used a
ratio of CO(4--3) to CO(1--0) of 0.55, as typically found for strongly
gravitationally lensed, dusty, far-infrared-selected high-redshift
galaxies \citep{spilker14,danielson11}. The difference compared with the 
lower estimate based on the CO(1--0) line detection of \citet{harrington18} 
is below 0.1~dex, and indicated by the small horizontal arrow in
Fig.~\ref{fig:sk}. Resulting gas-mass surface densities are
2400--$2900$~M$_{\odot}$~pc$^{-2}$.

\subsubsection{Star-formation law}

In Figure~\ref{fig:sk} we show that the Emerald falls just below 
the relationship for high-redshift starburst galaxies. Filled red 
and empty blue stars indicate the central surface brightnesses of 
the four bright clumps shown in Fig.~\ref{fig:comaps}, and the 
residual intraclump emission, respectively (see 
Section~\ref{ssec:clumpstability}). Star-formation and gas mass 
surface densities in the Emerald are comparable to submillimeter 
galaxies in the field \citep[][]{bothwell10}, and are significantly 
lower than in the Ruby \citepalias[][]{canameras17b}, SDP.81 
\citep[][]{hatsukade15}, and the Eyelash \citep[][]{swinbank11}. This 
includes the bright star-forming clumps, but also the extended diffuse 
emission, which has molecular gas mass surface densities of only a 
few~100 M$_{\odot}$~pc$^{-2}$.

It is important to test if the lower surface densities are a consequence of 
our larger beam sizes. The Ruby has clump sizes of 0.1\arcsec--0.3\arcsec, 
and if we run a toy model with clumps of such sizes, we see 
indeed a decrease in gas mass and star-formation surface density of a 
few 0.1~dex compared to the 0.1\arcsec\ beam with which the Ruby was 
observed \citepalias[][]{canameras17b}, although the position relative 
to the ridge lines of starburst and main-sequence galaxies does not
change. However, clumps as compact as those in the Ruby should have
clearly been seen in our SMA VEXT imaging. We ran a suite of simple
toy models to investigate how many sources with 0.2\arcsec\ FWHM size,
comparable to the clump sizes in the Ruby, would be necessary to
explain the 6-mJy minimal brightness we observed with the EXT
configuration of the SMA, without violating the 3$\sigma$ upper
surface-brightness limit imposed by the non-detection of such clumps
in the VEXT configuration down to rms~=~1.75~mJy. This was only possible
by populating the 0.8\arcsec\ beam with a near-uniform distribution of
more compact sources of about the same brightness. This strongly
disfavors the presence of bright, compact clumps within the Emerald
with sizes much less than the 0.6\arcsec\ minor axis size seen in
stellar light. Beam-smearing effects in Fig.~\ref{fig:sk} must
therefore be small, which implies star-formation intensities of a few 
tens of M$_{\odot}$~yr$^{-1}$~kpc$^{-2}$ and gas densities of a few thousand 
M$_{\odot}$~pc$^{-2}$.

\section{Feedback and disk fragmentation}
\label{sec:gasenergetics}

\subsection{Global disk (Toomre) stability}
\label{ssec:diskstability}

Most galaxies observed at redshifts $z \ga 2$ with resolved data sets
and discernable velocity gradients fall near the critical value for
rotationally supported gas, meaning that they have Toomre parameters $Q \sim 1$. 
For thin, uniform disk models, this implies that their gas reservoirs are 
globally stable against gravitational collapse on kpc scales. Several 
authors have recently pointed out that gas-rich, clumpy, star-forming 
galaxies that are already fragmented, may represent more complex 
environments where gas at $Q > 1$ can still form clumps. This pushes 
the critical stability parameter below which the gas becomes unstable 
to gravitational collapse to somewhat higher values, of order 
$Q \sim 2$--3 \citep[e.g.,][]{inoue16}.

\begin{figure*}
\centering
\includegraphics[width=0.3\textwidth]{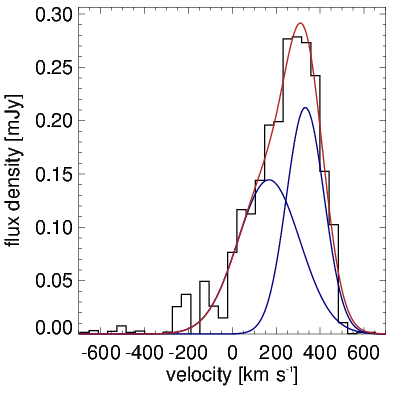}
\includegraphics[width=0.3\textwidth]{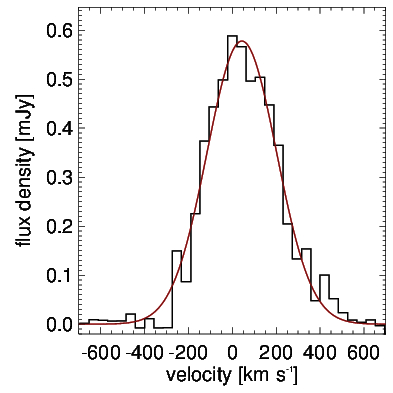}
\caption{Spectra of individual clumps \#1.2~S ({\it left}) and \#1.2~N ({\it right}),
extracted from our PdBI CO(4--3) data cube using 0.6\arcsec~$\times$~0.6\arcsec\ 
apertures, and corrected for gravitational magnification. Clump \#1.2~N 
is well fitted with a single Gaussian function (red curve), while \#1.2~S 
exhibits a secondary component, blueshifted with respect to the systemic 
component (blue curves), which we interpret as the signature of a stellar 
wind.}
\label{fig:clumpspec} 
\end{figure*}

The Emerald is no exception in this regard. Using $v_{\rm c}=380$~km
s$^{-1}$, $M_{\rm tot}=9.1 \times 10^{10}$~M$_{\odot}$, and
$M_{\rm gas}=1.1 \times 10^{10}$~M$_{\odot}$, and setting $Q= \sigma_0 / 
v_{\rm c} \times a \times M_{\rm tot} / M_{\rm gas}$ \citep[][]{genzel11}, we 
find $Q=1.3$. For this estimate we adopted the lowest value of $\sigma_0$
that we could find within the Emerald, $\sigma_0=95$~km~s$^{-1}$, which
is probably the best approximation of the gas kinematics outside the 
brightest star-forming knots, as suggested by \citet{inoue16}. Here $a$ 
is a morphological parameter which is of order unity \citep[see][for 
details]{genzel11}. The value of $Q=1.3$ is well within the range typically
found for high-redshift galaxies. Even the combination of estimates of
$v_{\rm c}$, $M_{\rm gas}$, $M_{\rm tot}$, and $\sigma$ that would yield the 
highest possible value of $Q$, would give $Q=2.6$, still in the range of
marginally Toomre-stable disks expected from simulations of fragmenting 
galaxies.

\subsection{Clump properties and stability}
\label{ssec:clumpstability}

Many authors have already discussed the importance of clump survival
for massive high-redshift galaxies \citep[e.g.,][]{cowie96,elmegreen07,
genzel08,bournaud09,fs09,tacconi13,mayer16,dessauges17}. Blue stellar 
continuum morphologies in about 50--60\% of actively star-forming, 
UV/optically-selected galaxies at high redshift show that considerable 
fractions of the stellar mass in these galaxies are in giant clumps of 
1~kpc or less in size and a few times 10$^7$ up to 10$^9$~M$_{\odot}$ 
in mass \citep[e.g.,][]{elmegreen05,forster11,livermore12}. Moreover 
disk fragmentation and clump formation can also be important for dusty, 
far-infrared and submillimeter selected galaxies like the GEMS 
\citep[][]{swinbank10}. If clumps are long-lived, they may sink toward
the galaxy center within a few orbital times (few hundred Myr) and 
merge to form a bulge. This scenario has been put forward in particular 
in the context of early simulations of gas-rich, fragmented disk 
galaxies \citep[e.g.,][]{ceverino10,bournaud14,mandelker14}, but has 
recently been challenged by detailed hydrodynamic models 
\citep[][]{tamburello15,mayer16,oklopcic17} and observations of blue,
gravitationally lensed galaxies \citep[][]{dessauges17,tamburello15}. 
These studies favor a scenario where clumps are more marginally 
gravitationally bound, and may dissolve within a few tens of Myr if 
feedback becomes too strong, either through mass loss in the form of winds,
or by producing turbulent velocities near or above the virial velocity, or 
a mixture of both \citep[for the latter, see in particular][]{hayward17}. 
Whether clumps survive seems to depend critically on the detail of how 
feedback is implemented in these simulations \citep[][]{mayer16}.

\begin{table*}
\centering
\begin{tiny}
\begin{tabular}{lccccccccccc}
\hline 
\hline
Source & RA & Dec & $\mu~S_{\rm 850}$ & SFR & $\mu$ & $I_{\rm CO(4-3)}$ & $M_{\rm gas}$ & $v$ & FWHM & $\alpha_{\rm vir}$ \\ 
 & (J2000) & (J2000) & [mJy] & [M$_{\odot}$ yr$^{-1}$] &  & [Jy km s$^{-1}$] & [$10^9\ M_{\odot}$] & [km s$^{-1}$] & [km s$^{-1}$]& \\
\hline 
1.2~N & 11:27:14.67 & $+$42:28:23.9 & 7.8 $\pm$ 0.6 & 100 $\pm$ 8 & 14.2 & $(2.3 \pm 0.2) \times 10^{-1}$ & 5.0 $\pm$ 0.5 & 44 $\pm$ 9 & 374 $\pm$ 38 & 2.9 $\pm$ 0.6 \\
1.2~S (m) & 11:27:14.70 & $+$42:28:23.0 & 8.1 $\pm$ 0.6 & 41 $\pm$ 3 & 35.0 & $(5.0 \pm 0.5) \times 10^{-2}$ & 1.0 $\pm$ 0.1 & 334 $\pm$ 28 & 199 $\pm$ 17 & 1.4 $\pm$ 0.3 \\
1.2~S (b) & -- & -- & -- & -- & -- & $(5.0 \pm 0.6) \times 10^{-2}$ & 1.2 $\pm$ 0.1 & 135 $\pm$ 17 & 333 $\pm$ 43 & 5.3 $\pm$ 1.1 \\
\hline
\end{tabular}
\end{tiny}
\caption{Properties of individual star-forming clumps in the Emerald. We list 
observed values for continuum flux densities, $S_{\rm 850}$, and intrinsic 
values for line fluxes, star-formation rates and gas masses. The local 
magnification factors are listed in column $\mu$. Error bars give the rms 
for the continuum and uncertainties in the fit for emission-line data. 
For the clumps where we identify a secondary line component, we also list 
the velocity offset and FWHM line width at the position of that clump. The 
virial parameters, $\alpha_{\rm vir}$, quantify the relative contribution 
from gravitational binding and turbulent energy.}
\label{tab:clumps}
\end{table*}

In the Emerald, the degree of fragmentation between the stellar
component on the one hand, and gas and dust on the other, is very
different. Stars, dust, and gas in the Emerald are found over large
and similar ranges in a disk of at least 2.7~kpc~$\times$~1.7~kpc in
size, and with rather moderate gas and star-formation rate surface
densities (Sect.~\ref{ssec:schmidtkennicutt}). The stellar light
distribution is very smooth. At most about 25\% of the F110W emission
from the arc is in high surface-brightness features, and less than a
few percent in the F160W image. Overall, the surface brightness varies 
by less than factors of between two and three in the F160W image, and 
varies only on small scales comparable to the PSF size, which would be 
smoothed out if seen at a resolution comparable to those of our 
long-wavelength data.

In contrast, most of the gas and dust emission is concentrated in two
bright clumps with radii $\la 100$--250~pc, each imaged to the north and 
south of the critical line (see Sect.~\ref{ssec:starformation} and 
Fig.~\ref{fig:comaps}). These are upper limits set by the beam size and 
local gravitational magnification factors listed in Table~\ref{tab:clumps}.
The two clumps together comprise about 60\% of the total dust emission
in the Emerald. A fainter, diffuse component is also seen after
modeling the clumps with a Gaussian beam, and removing their contribution 
to the emission. Moreover, the average stellar mass surface density of 
$1.1 \times 10^{10}$~M$_{\odot}$~kpc$^{-2}$ is a factor of a few greater than 
those of the molecular gas inferred in Sect.~\ref{ssec:schmidtkennicutt} 
(see Fig.~\ref{fig:sk}). This results in a global gas-to-baryonic mass 
ratio in the Emerald of about 20\%.

We show the spectra of the two clumps in Fig.~\ref{fig:clumpspec}, 
extracted over $3 \times 3$~pixel apertures in our CO line data
(0.6\arcsec~$\times$~0.6\arcsec), and scaled to the flux density per 
beam of the central pixel, as is most appropriate for unresolved sources.
We have focused on the two northern images of these clumps, \#1.2~N and \#1.2~S, 
which are easier to deblend. Their properties are derived following the 
same approach and assumptions used for the Schmidt-Kennicutt analysis in 
Section~\ref{ssec:schmidtkennicutt}. Both clumps are massive and gas-rich, 
with $M_{\rm gas}=5.0 \times 10^9$~M$_{\odot}$ and $1.0\times 10^9$~M$_{\odot}$, 
respectively, in the upper range of clump masses in other 
high-redshift galaxies with giant clumps \citep[][]{genzel11}. 
Table~\ref{tab:clumps} lists their individual properties, including 
star-formation rates of 40--100~M$_{\odot}$~yr$^{-1}$ (for a Chabrier IMF), 
and FWHM line widths of CO(4--3) of 200--375~km~s$^{-1}$. Their sizes are 
smaller than those seen in field galaxies with very massive clumps \citep[][but 
we note that these masses may be overestimated, \citealt{tamburello15,
dessauges17}]{forster11}. Gas mass and star-formation rate surface 
densities are also similar to both massive high-redshift clumps and 
submillimeter galaxies (Fig.~\ref{fig:sk}).

We follow \citet{oklopcic17} in using the virial parameter, 
$\alpha_{\rm vir}$, to investigate whether these clumps are gravitationally 
bound. The quantity $\alpha_{\rm vir}$ was first introduced by 
\citet{bertoldi92} and can be used as a measure to compare
gravitational binding and turbulent energy in the gas, $\alpha_{\rm vir}=
5 \sigma^2 R / G M_{\rm gas}$, where $\sigma$ is the velocity dispersion
of the main emission line components of the clumps, $R$ the clump
radius, $M_{\rm gas}$ the gas mass in this emission line, and $G$ the 
gravitational constant. Values of $\alpha_{\rm vir} \sim 1$--5 are typical 
for star-forming molecular gas at low redshift \citep[][]{heyer09}, and
have also been observed at very high star-formation intensities at
high redshift \citepalias[][]{canameras17b}. They are considered a
signature of turbulence-regulated star formation \citep[][]{krumholz05}. 
Values below unity show that most of the gas within the structure is 
likely gravitationally bound.

With the radius set by the beam size and gravitational magnification
at the cloud center, and cloud mass and velocity dispersion as listed
in Table~\ref{tab:clumps}, we find values for $\alpha_{\rm vir}$ of 1.4 
and 2.9 for the two systemic components. $\alpha_{\rm vir}$ would of course 
be lower if we had strongly overestimated $R$, as would seem possible for 
spatially unresolved clouds. However, this would be in contradiction with 
the non-detections of these clumps with the SMA in the VEXT configuration, 
which rule out very compact sizes with projected diameters much below 
about~0.6\arcsec\ (Sect.~\ref{ssec:schmidtkennicutt}).

Our results suggest that the gas in these clumps is marginally
gravitationally bound, as in many star-forming molecular clouds in the
Milky Way, and perhaps also in intensely star-forming galaxies at high
redshift out to the highest gas mass and star-formation surface 
densities \citepalias[][]{canameras17b}. This is consistent with the
simulations of \citet{mayer16} and \citet{oklopcic17} with detailed 
feedback descriptions. These authors argue that feedback in clumpy 
high-redshift galaxies with stellar masses and gas fractions comparable 
to the Emerald should maintain star-forming clumps near or above 
$\alpha_{\rm vir} \sim 1$. It is possible that this is due to the
relatively low global gas fraction in the Emerald of 20\%
\citep[][]{bournaud14,renaud18}, although the example of the Ruby, one
of the most intensely star-forming galaxies at high redshift, also
suggests $\alpha_{\rm vir}\ga 1$ \citepalias[][]{canameras17b}. The two 
clumps we consider here have masses of about $10^9$~M$_{\odot}$, in the 
upper mass range in these simulations. We now analyze our evidence 
for feedback in the Emerald, before discussing its potential role for 
the clumps and disk in this galaxy.

\subsection{Feedback from star formation}
\label{ssec:feedback}

Several authors have recently discussed the importance of winds for cloud 
stability \citep{genel12,mayer16,oklopcic17}. If winds remove large fractions 
of the cloud mass in a few tens of Myr, they may lead to clump dissolution, 
even if the initial mass was large enough to form a self-gravitating and 
star-forming cloud. These timescales are shorter than the age of the young 
stellar population in the Emerald, and the smooth distribution of stellar 
light compared to the clumpy star formation may provide circumstantial 
evidence for this.

The spectra in Fig.~\ref{fig:clumpspec} show that only one clump, \#1.2~N, 
can be fitted well with a single Gaussian distribution, the other, \#1.2~S, 
has a pronounced blue component with FWHM~$=333$~km~s$^{-1}$ and a blueshift 
of $-199$~km~s$^{-1}$ relative to the systemic line, which is also seen as a 
secondary component in the line maps in Figure~\ref{fig:comaps}. Blueshifted 
emission-line components are generally viewed as a characteristics of winds, 
which may or may not escape from the host galaxy, depending on the outflow 
velocity and depth of the gravitational potential. While emission-line 
signatures of winds of warm ionized gas with velocities of a few hundred 
km~s$^{-1}$ are common at high redshift \citep[e.g.,][]{letiran11,nesvadba07,
genzel11}, this is, to our knowledge, the first such component seen in
molecular gas in an intensely star-forming galaxy at high redshift. An
example for a similar, but more extreme component in the CO line emission 
of a high-redshift quasar has been given by \citet{nesvadba11}. At low 
redshift, multiple examples of wind components seen in molecular gas exist 
\citep[e.g.,][]{weiss99,sturm11,walter17}. Observing a blueshifted component
may imply that we only see one side of the wind, which is lifted off an
optically thick gas disk. Given that the region very near the critical
line also has the most prominent stellar continuum emission, it is 
possible that the wind has lowered the overlying columns of dust and gas. 
We now analyze the physical properties of this outflow in more detail, 
and discuss its likely impact for the clump from which it originates, and
the galaxy overall.

\subsubsection{Kinetic energy and momentum estimates}

\begin{table*}
\centering
\begin{tiny}
\begin{tabular}{lcccccccccccccc}
\hline \hline
Source & $\Delta v$ & $E_{\rm \Delta v}$ & $P_{\rm \Delta v}$ & $E_{\sigma}$ & $P_{\sigma}$ & $dE_{\rm SF}/dt$ & $dP_{\rm SF}/dt$ \\ 
 & [km s$^{-1}$] & [$10^{56}$ erg] & [$10^{49}$ dyn s] & [$10^{56}$ erg] & [$10^{49}$ dyn s] & [$10^{43}$ erg s$^{-1}$] & [$10^{35}$ dyn] \\
\hline
1.2~S (b) & 199 $\pm$ 28 & 4.7 & 2.4 & 7.1 & 5.0 & 2.6 & 2.0 \\
\hline
\end{tabular}
\end{tiny}
\caption{Properties of the molecular wind toward star-forming clump \#1.2~S. 
The columns are: velocity offset between the wind and systemic component,
$\Delta v$; energy ($E_{\rm \Delta v}$) and momentum ($P_{\rm \Delta v}$) 
derived from the observed velocity offsets; energy ($E_{\sigma}$) and 
momentum ($P_{\sigma}$) corresponding to the observed line widths; energy
and momentum injection rates, $dE_{\rm SF}/dt$ and $dP_{\rm SF}/dt$, from
star formation. Momentum and energy injection rates are normalized to 
an outflow timescale of 50~Myr. We provide a measurement uncertainty
only for the velocity offset. For the other estimates, typical 
measurement uncertainties are between 10 and 20\%, and likely much 
smaller than the astrophysical uncertainties, which are difficult to 
quantify accurately.}
\label{tab:clumpwinds} 
\end{table*}

With the same assumptions as in Sect.~\ref{ssec:schmidtkennicutt}, we
estimated an intrinsic molecular gas mass of $1.2\times 10^{9}$~M$_{\odot}$ 
for the secondary component in clump \#1.2~S. With this mass estimate, 
and following our earlier analysis of the Ruby \citepalias[][]{canameras17b}, 
we were able to use the kinetic energy and momentum in this component as 
constraints to investigate whether the star formation is powerful enough to 
produce a wind with the observed properties.

We followed \citet{heckman15} to estimate the momentum, and
\citetalias{canameras17b} to estimate the energy injection rates from 
star formation into the gas. We discarded a contribution from an AGN, 
because the spectral energy distribution shows no evidence of 
one \citepalias[][]{canameras15}. \citet{heckman15} showed that starbursts 
may inject $(4.8\times 10^{33}) \times {\rm SFR}$~dyn of momentum flux
into the gas per unit of stellar mass formed, where the star-formation
rate is given in M$_{\odot}$~yr$^{-1}$. This estimate is matched
to observed outflows in very vigorous low-redshift starbursts, and is
also in the range of long-term winds proposed by \citet{dekel13} for
giant clumps. It includes contributions from radiation pressure as
well as mechanical feedback and is valid for gas entrained in a hot
wind, following the \citet{chevalier85} approach. For the
41~M$_{\odot}$~yr$^{-1}$ of star formation in \#1.2~S, this corresponds 
to a momentum injection rate of $2 \times 10^{35}$~dyn. Corresponding 
values for the other clump and the Emerald overall are $5.0$ and 
$8.4\times 10^{35}$~dyn, respectively, for star-formation rates of 
100 and 176~M$_{\odot}$ yr$^{-1}$.

As in \citetalias{canameras17b}, we relied on Starburst99 
\citep[][]{leitherer99} to estimate that continuous star formation over a
few times $10^7$~yrs, solar metallicity and a Chabrier stellar IMF would 
produce a mechanical luminosity of $10^{41.8}$~erg s$^{-1}$ for each newly 
formed solar mass. For SFR $=41$~M$_{\odot}$~yr$^{-1}$, this corresponds to 
$2.6\times 10^{43}$~erg s$^{-1}$ in clump \#1.2~S, and to 6 and 
$11 \times 10^{43}$~erg s$^{-1}$ for star-formation rates of 100 and 
176~M$_{\odot}$~yr$^{-1}$ in clump \#1.2~N and the Emerald, respectively. 
All results obtained from clump \#1.2~S are also listed in 
Table~\ref{tab:clumpwinds}.

Following our earlier analysis of the Ruby \citepalias[][]{canameras17b},
we were able to constrain the kinetic energy and momentum of the gas. 
For the kinetic energy we set $E_{\Delta v}=1/2\ M\ v^2$ for bulk, and
$E_{\sigma}=3/2\ M\ \sigma^2$ for unordered motion, and we derived the
corresponding momentum by setting $P_{\rm wind}=E_{\rm wind}/v$ 
(see Table~\ref{tab:clumpwinds}). With 
$\sigma_{\rm wind}=141$~km~s$^{-1}$ and $\Delta v=-199$~km~s$^{-1}$
velocity offset between blueshifted and systemic component, we find 
an average combined kinetic energy of $11.8 \times 10^{56}$~erg in the
outflowing gas, and a momentum of $7.4 \times 10^{49}$~dyn~s. These
kinematics can be powered by the starburst in clump \#1.2~S if 
the current momentum and energy injection rates are being maintained 
for at least 11.7~Myr and 1.5~Myr, respectively. This is less than the 
age of the starburst of about 50~Myr estimated in 
Sect.~\ref{ssec:stellarpop}.

\subsubsection{Discussion}

Is the starburst in the Emerald powerful enough to unbind the gas
from the clump, and from the galaxy overall? To test the first question, 
we assumed that the clump is a virialized gas sphere, and set
$v_{\rm esc}=\sqrt{2} v_{\rm vir}= \sqrt{2 M\ G/5\ R}$ = 97~km~s$^{-1}$ for a
clump with mass $M$ and radius $R$ as observed. $v_{\rm esc}$ and $v_{\rm vir}$ 
are the escape and virial velocity, respectively. The resulting escape 
velocity is significantly lower than the velocity offset of 199~km~s$^{-1}$ 
found between the wind and systemic component (even while discarding 
projection effects), suggesting that much of the outflowing gas will 
escape from the clump.

However, the same does not necessarily hold for the galaxy overall. 
For a galaxy with at least $5\times 10^{10}$~M$_{\odot}$ of mass, as 
implied by our gas and stellar mass estimates, we would expect an escape 
velocity, $v_{\rm esc}$, of at least $v_{\rm esc}= \sqrt{2}\ v_c$, i.e., 
380~km~s$^{-1}$ for a disk with $R=3$~kpc, compared to a velocity offset 
of 199~km s$^{-1}$. The same is suggested by a more detailed estimate 
following \citet{ostriker11} and \citet{newman12}. We assumed that the
wind is mainly momentum-driven, and set ($P_{\rm tot}/dM_*/dt) = 7.2 \times
f_{\rm g,0.1} \times \Sigma_{\rm d,1000}^2 / \Sigma_{\rm SFR,100}$, where
$(P_{\rm tot}/dM_*/dt)$ is the characteristic momentum injection rate or,
in other words, the total momentum, $P_{\rm tot}$, injected by the star formation
per unit mass formed, $dM_*/dt$. The quantity $f_{\rm g,0.1}$ is the gas 
fraction in units of 0.1, $\Sigma_{\rm d,1000}^2$ is the disk mass surface 
density in units of 1000~M$_{\odot}$~pc$^{-2}$, and $\Sigma_{\rm SFR,100}$ 
is the star-formation rate surface density in units of 100~M$_{\odot}$
yr$^{-1}$ kpc$^{-2}$. With a gas-to-baryonic mass fraction of 0.2, disk mass
surface density $1.3\times 10^{10}$ M$_{\odot}$ kpc$^{-2}$, and
$\Sigma_{\rm SFR}=49$ M$_{\odot}$ yr$^{-1}$ kpc$^{-2}$ (for ${\rm SFR}=176$
M$_{\odot}$ yr$^{-1}$ and a disk surface of 3.6~kpc$^2$), we find that
each solar mass worth of stars formed must provide a characteristic
momentum injection rate of 4990 km s$^{-1}$ to balance the hydrostatic
mid-plane pressure of the galaxy. This value is much greater than
expected for purely radiation-pressure driven winds \citep[1000 km
s$^{-1}$,][]{murray05}, and also for winds including the momentum
injection from supernovae and young stars \citep[2500--3000 km
s$^{-1}$,][]{ostriker11,heckman15}. It is therefore unlikely that 
this wind is very efficient in removing gas from the Emerald.

Nonetheless, the wind has a potentially important impact on the future 
of the clump. Assuming that the gas in both components of clump \#1.2~S
can be modeled with the same CO-to-H$_2$ conversion factor, we find a
mass-outflow rate of $dM/dt=247$~M$_{\odot}$~yr$^{-1}$, for a dynamical 
time estimate $t_{\rm dyn} = R/v=1.5$~Myr, using $R=300$~pc and 
$v=199$~km~s$^{-1}$. This corresponds to a mass-loading factor of 
about five, not unusually high for a momentum-driven wind, and suggests 
that the clump may lose most of its mass in about 2~Myr. This is
comparable to the free-fall time of gas with average density of a few
times 100~cm$^{-3}$, in the range of the average density of a gas 
sphere with a mass of about $5 \times 10^8$~M$_{\odot}$ and radius of 
100~pc. While this is a highly simplified toy model and neglects any
contribution from gas accretion onto the clump \citep[which can be
considerable, e.g.,][]{dekel13}, it does highlight that these outflow
times are very short, and that these clumps are likely to be transient
structures whose survival depends sensitively on the dynamical
equilibrium between gas accretion and outflows, as suggested
previously by simulations, and as also found for giant molecular clouds 
in the Milky Way \citep[e.g.,][]{murray11}.

Although much of the outflowing gas is likely to escape from the clump
itself, it remains gravitationally bound to the host galaxy, and will
therefore ultimately fall back and be available for star formation
again. Galactic fountains are well studied in the nearby Universe
\citep[e.g.,][]{marinacci11,sancisi08}, although their re-accretion
times are very long compared to the relevant timescales in these
clumps. However, given the much higher gas densities and accordingly
shorter cooling times in our case, it is possible that the fall-back 
timescales are also much shorter. It may be that re-accretion of gas 
from such fountains contributes significantly to the accretion that is 
required to maintain the clumps marginally bound and star-forming over 
more extended periods of time, and potentially also to the turbulence 
observed in the Emerald. A detailed test of this scenario would 
require hydrodynamic modeling.

\section{Summary and conclusions}
\label{sec:summary}

We have presented a detailed study of the molecular gas, dust and 
stellar components in the Emerald (PLCK\_G165.7+49.8), as observed 
with IRAM and SMA interferometry of the CO(4--3) line and 850~$\mu$m dust 
emission, and {\it HST}/WFC3 and CFHT optical and near-infrared imaging. 
The Emerald is a strongly gravitationally lensed dusty star-forming 
galaxy at $z=2.236$, part of \textit{Planck}'s Dusty GEMS and surrounded 
by two compact submillimeter sources and multiple extended, 
near-infrared gravitational arcs. It falls behind a previously unknown
massive galaxy cluster and associated large-scale filament at
$z=0.348$. We characterized the foreground environment through the
density distribution, color-magnitude diagram, and photometric and
spectroscopic redshifts of member galaxies. We used the cluster
members and eight multiply imaged systems of background sources to
constrain the strong lensing mass distribution with {\sc Lenstool}, 
and computed the magnification factors of the Emerald. The Emerald is
composed of a main arc formed by two images of the same region in the
source plane, with intrinsic size of about 2.7~kpc~$\times$~1.7~kpc,
and magnification factors of $\mu_{\rm dust}= 29.4 \pm 5.9$, $\mu_{\rm
gas}= 24.1 \pm 4.8$ and $\mu_{\rm stars}= 34.1 \pm 6.8$ for the dust,
gas, and stars. The two nearby compact submillimeter sources are
magnified by more-moderate factors. One is likely an additional 
counter-image of the source producing the Emerald, the other is a 
separate galaxy at the same redshift. With 18~kpc projected distance 
it is probably a companion galaxy of the Emerald.

The Emerald has intrinsic properties of a typical ULIRG, with a 
far-infrared luminosity, $L_{\rm FIR}= (1.8 \pm 0.4) \times 10^{12}$~L$_{\odot}$, 
SFR of $(176 \pm 35)$~M$_{\odot}$~yr$^{-1}$, and dust and molecular gas 
masses of $(7.4 \pm 1.5) \times 10^7$~M$_{\odot}$ and
$(1.1 \pm 0.2) \times 10^{10}$~M$_{\odot}$, respectively. Gas and dust
morphologies are clumpy, and the gas kinematics exhibit strong
velocity offsets, symmetric on both sides of the arc due to the parity
flip at the position of the critical curve. Interpreting the
double-imaged velocity gradient of 380~km~s$^{-1}$ over the arc as the
partially-sampled rotation curve of a gaseous disk, we obtain a
dynamical mass of $9.1 \times 10^{10}$~M$_{\odot}$. Gaussian line widths
are in the range 90--190~km~s$^{-1}$, consistent with a marginally
stable disk with a Toomre parameter $Q \sim 1.3$.

The stellar morphology seen with \textit{HST}/WFC3 is diffuse and 
filamentary, in contrast to the dust and gas morphologies, which are 
dominated by two doubly imaged clumps. The total, intrinsic stellar 
mass and mass surface density are $(4.1 \pm 0.4) \times 10^{10}$~M$_{\odot}$ 
and $(11 \pm 5) \times 10^9$~M$_{\odot}$~kpc$^{-2}$, respectively, similar 
to other high-redshift dusty starburst galaxies. We also find 
$A_{\rm V}=3.9 \pm 0.3$~mag and an average gas-to-baryon fraction of
about 20\%.

We probed the resolved Schmidt-Kennicutt law in the two clumps, finding
star-formation intensities of about 40~M$_{\odot}$ yr$^{-1}$ kpc$^{-2}$,
placing them just below the relationship for high-redshift starburst
galaxies, and near to the properties of other well-studied high-redshift 
clumps and dusty star-forming galaxies in the literature. The more 
diffuse intraclump emission with gas mass surface densities down to
600~M$_{\odot}$~pc$^{-2}$ falls into a similar regime.

The two clumps are massive, with gas masses of 1.0 and $5.0 \times
10^{9}$~M$_{\odot}$, and star-formation rates of 40 and 
100~M$_{\odot}$~yr$^{-1}$, respectively, and upper limits on their 
FWHM sizes of 200--500~pc. They are marginally gravitationally bound. The 
integrated CO(4--3) line profile of one of these clumps shows a significant 
blueshifted wing, offset by about $-200$~km~s$^{-1}$ relative to the main
emission-line component of the clump. We interpret this offset in the
usual way, that is, as an outflow signature. To our knowledge, this is
the first signature of a molecular wind arising from a massive clump 
in a dusty star-forming galaxy
at high redshift. The kinetic energy and momentum injection rates from
star formation in the nearby massive clump are sufficient to drive
such a wind, and to make the gas escape from the clump, but probably
not from the host galaxy itself. This molecular wind has a higher
mass-loading factor than those measured for ionized winds in
$z \sim 2$ galaxies, as expected in scenarios where clumps are
short-lived and dissolve on timescales of a few tens of Myr, except 
if most of the gas is replenished by continuous accretion onto the 
clump. Unless we see the outflow at an extreme inclination angle, the 
gas will probably remain bound to the host galaxy and will form a 
galactic fountain, potentially contributing to the fueling of subsequent 
star formation in the Emerald.

\section*{Acknowledgments}

The authors would like to thank the anonymous referee for providing 
useful comments that helped improve the paper. RC would like to 
thank Raphael Gavazzi, Claudio Grillo, Lise Christensen and Vera 
Patr\'icio for useful discussions. We acknowledge the staff at IRAM 
and SMA for carrying out the observations on which this program is 
based. A major part of this work is based on observations carried 
out with the Plateau de Bure Interferometer of IRAM, and the 
Canada-France-Hawaii Telescope and the Submillimeter Array on top of 
Mauna Kea, Hawaii. IRAM is supported by INSU/CNRS (France), MPG 
(Germany) and IGN (Spain). The Submillimeter Array is a joint project 
between the Smithsonian Astrophysical Observatory and the Academia 
Sinica Institute of Astronomy and Astrophysics and is funded by the 
Smithsonian Institution and the Academia Sinica. The Canada-France-Hawaii 
Telescope (CFHT) which is operated by the National Research Council of 
Canada, the Institut National des Sciences de l'Univers of the Centre 
National de la Recherche Scientifique of France, and the University of 
Hawaii. This work is also based in part on observations made with the 
Spitzer Space Telescope, which is operated by the Jet Propulsion 
Laboratory, California Institute of Technology under a contract with NASA. 
This publication makes use of data products from the Two Micron All Sky 
Survey, which is a joint project of the University of Massachusetts and 
the Infrared Processing and Analysis Center/California Institute of 
Technology, funded by the National Aeronautics and Space Administration 
and the National Science Foundation. It also uses data from SDSS-III,
funded by the Alfred P. Sloan Foundation, the Participating Institutions, 
the National Science Foundation, and the U.S. Department of Energy Office 
of Science. The SDSS-III web site is http://www.sdss3.org/. This work 
was supported by the Programme National Cosmology et Galaxies (PNCG) of 
CNRS/INSU with INP and IN2P3, co-funded by CEA and CNES. RC is supported 
by DFF -- 4090-00079. ML acknowledges CNRS and CNES for support.

\bibliography{g165}

\end{document}